\pdfoutput=1
\newif\iflipics\lipicstrue
\newif\ifappendix\appendixtrue
\documentclass{lmcs}
\pdfoutput=1

\usepackage{lastpage}
\lmcsdoi{15}{4}{18}
\lmcsheading{}{\pageref{LastPage}}{}{}%
{Mar.~16,~2018}{Dec.~19,~2019}{}

\keywords{probabilistic, relational lifting, differential privacy, max flow/min cut}

\usepackage[utf8]{inputenc}
\usepackage[T1]{fontenc}
\usepackage{microtype}

\usepackage{etex,etoolbox,environ,amsthm}

\makeatletter
\newenvironment{xproof}[1][\proofname]{\par
  \begin{proof}%
}{%
  \end{proof}%
}

\providecommand{\@fourthoffour}[4]{#4}
\def\fixstatement#1{%
  \AtEndEnvironment{#1}{%
    \xdef\pat@label{\expandafter\expandafter\expandafter
      \@fourthoffour\csname#1\endcsname\space\@currentlabel}}}

\globtoksblk\prooftoks{1000}
\newcounter{proofcount}

\long\def\proofatend#1\endproofatend{%
  \refstepcounter{proofcount}%
  \begin{xproof} See Appendix, p.~\pageref{proof:\roman{proofcount}} \end{xproof}%
  \edef\next{\noexpand\begin{proof}[Proof of \pat@label]\noexpand\label{proof:\roman{proofcount}}}%
  \toks\numexpr\prooftoks+\value{proofcount}\relax=\expandafter{\next#1\end{proof}}}

\long\def\xproofatend#1\xendproofatend{%
  \refstepcounter{proofcount}%
  \edef\next{\noexpand\begin{proof}[Proof of \pat@label]}%
  \toks\numexpr\prooftoks+\value{proofcount}\relax=\expandafter{\next#1\end{proof}}}

\def\printproofs{%
  \count@=\z@
  \loop
    \the\toks\numexpr\prooftoks+\count@\relax
    \ifnum\count@<\value{proofcount}%
    \advance\count@\@ne
  \repeat}
\makeatother

\fixstatement{thm}
\fixstatement{lem}

\ifappendix\else
\long\def\proofatend#1\endproofatend{}
\long\def\xproofatend#1\xendproofatend{}
\fi

\usepackage[only,llbracket,rrbracket]{stmaryrd}
\SetSymbolFont{stmry}{bold}{U}{stmry}{m}{n}

\usepackage{amsmath,amsthm,amssymb,mathtools,nicefrac}
\usepackage[bb=boondox]{mathalfa}
\usepackage{xspace}
\usepackage{enumitem}
\usepackage{bbding}
\usepackage[sort&compress]{natbib}

\makeatletter
\def\NAT@spacechar{~}
\makeatother

\usepackage{todonotes}
\usepackage{setspace}

\newcounter{todocnt}

\def\ie{i.e.\xspace}
\def\eg{e.g.\xspace}

\newcounter{localc}

\newcommand{\eqdef}{\mathrel{\stackrel{\scriptscriptstyle \triangle}{=}}}
\newcommand{\iffdef}{\mathrel{\stackrel{\scriptscriptstyle \triangle}{\iff}}}

\newcommand{\inv}[1]{#1^{\raisebox{.2ex}{$\scriptscriptstyle-\!1$}}}

\newcommand{\proj}[1]{\pi_{#1}}
\newcommand{\fst}{\proj{1}}
\newcommand{\snd}{\proj{2}}

\newcommand{\rR}{\mathrel{\mathcal{R}}}
\newcommand{\rS}{\mathrel{\mathcal{S}}}

\newcommand{\oR}{\mathcal{R}}
\newcommand{\oS}{\mathcal{S}}

\newcommand{\rcomp}[1]{\overline{#1}}

\def\ssrc{\mathop{\top}}
\def\sdst{\mathop{\perp}}

\newcommand{\cset}[1]{\mathcal{E}(#1)}

\DeclareFontFamily{U}{mathx}{\hyphenchar\font45}
\DeclareFontShape{U}{mathx}{m}{n}{
      <5> <6> <7> <8> <9> <10>
      <10.95> <12> <14.4> <17.28> <20.74> <24.88>
      mathx10
      }{}
\DeclareSymbolFont{mathx}{U}{mathx}{m}{n}
\DeclareMathSymbol{\bigtimes}{1}{mathx}{"91}

\newcommand{\rset}[1]{\ensuremath{\mathbb{#1}}}

\newcommand{\BB}{{\{0, 1\}}}
\newcommand{\NN}{\rset{N}}
\newcommand{\ZZ}{\rset{Z}}

\newcommand{\RR}{\rset{R}}

\renewcommand{\setminus}{\mathrel{-}}

\newcommand{\DistOp}{{\mathbb{D}}}

\newcommand{\Exp}{\mathbb{E}}
\newcommand{\ExpD}{\mathbb{E}}
\newcommand{\PrS}{\mathbb{P}}

\newcommand{\Dist}{\DistOp}

\newcommand{\dnull}[1][]{{{\mathbb{0}}^{#1}}}
\newcommand{\dunit}[2][]{{{\mathbb{1}}^{#1}_{#2}}}

\newcommand{\dlet}[3]{\ExpD_{{#1} \sim {#2}} [{#3}]}

\newcommand{\dlift}[1]{#1^{\sharp}}
\newcommand{\dproj}[1]{\dlift{\proj{#1}}}
\newcommand{\dfst}{\dproj{1}}
\newcommand{\dsnd}{\dproj{2}}
\newcommand{\drestr}[2]{{#1}_{|{#2}}}

\newcommand{\sE}[2]{\Exp_{{#1}} [{#2}]}

\renewcommand{\P}[3]{\PrS_{{#1} \sim {#2}} [{#3}]}
\newcommand{\sP}[2]{\PrS_{#1} [{#2}]}

\newcommand{\mass}[1]{|{#1}|}
\DeclareMathOperator{\supp}{supp}

\newcommand{\dalifted}[7]
  {\langle {#1, #2} \rangle \mathrel{\blacktriangleleft_{#3,#4}^{#5}}
     \langle {#6} \mathrel{\&} {#7} \rangle}

\newcommand{\pre}{\phi}

\def\ms{\mspace{-1.5mu}}

\newcommand{\lmark}{{\scriptscriptstyle \vartriangleleft}}
\newcommand{\rmark}{{\scriptscriptstyle \vartriangleright}}

\newcommand{\lside}{_{\ms\lmark}}
\newcommand{\rside}{_{\ms\rmark}}

\newcommand{\lift}[1]{\mathrel{#1^\sharp}}
\newcommand{\alifttoplas}[2]{\mathrel{#1^{(1)}_{#2}}}
\newcommand{\alifticalp}[2]{\mathrel{#1^{(2)}_{#2}}}
\newcommand{\aliftnew}[2]{\mathrel{#1^{(\star)}_{#2}}}

\newcommand{\symlifttoplas}[2]{\mathrel{\overline{#1}^{(1)}_{#2}}}
\newcommand{\symliftnew}[2]{\mathrel{\overline{#1}^{(\star)}_{#2}}}

\usepackage{hyperref}

\iflipics
\else
\usepackage[capitalise]{cleveref}

\crefname{section}{section}{sections}
\Crefname{section}{Section}{Sections}

\crefname{lemma}{lemma}{lemmas}
\Crefname{lemma}{Lemma}{Lemmas}

\crefname{theorem}{theorem}{theorems}
\Crefname{theorem}{Theorem}{Theorems}

\crefname{definition}{definition}{definitions}
\Crefname{definition}{Definition}{Definitions}
\fi

\renewcommand{\epsilon}{\varepsilon}

\usepackage{relsize,tikz}
\usetikzlibrary{matrix,arrows,backgrounds,calc,shapes,fit}
\usetikzlibrary{decorations.pathmorphing}

\tikzset{
    between/.style args={#1 and #2}{
         at = ($(#1)!0.5!(#2)$)
    }
}

\tikzset{fontscale/.style = {font=\relsize{#1}}}

\begin{document}

\title{Relational \texorpdfstring{$\star$-Liftings}{*-Liftings} for Differential Privacy}

\author[G. Barthe]{Gilles Barthe}
\address{IMDEA Software Institute, Spain and MPI for Security and Privacy, Germany}
\email{gilles.barthe@imdea.org}

\author[T. Espitau]{Thomas Espitau}
\address{Sorbonne Universités, UPMC Paris 6, France}
\email{t.espitau@gmail.com}

\author[J. Hsu]{Justin Hsu}
\address{University of Wisconsin--Madison, USA}
\email{email@justinh.su}

\author[T. Sato]{Tetsuya Sato}
\address{Seikei University, Japan}
\email{t\_sato@st.seikei.ac.jp}

\author[P.-Y. Strub]{Pierre-Yves Strub}
\address{École Polytechnique, France}
\email{pierre-yves@strub.nu}

\begin{abstract}
  Recent developments in formal verification have identified \emph{approximate
    liftings} (also known as \emph{approximate couplings}) as a clean,
  compositional abstraction for proving differential privacy. This construction
  can be defined in two styles. Earlier definitions require the
  existence of one or more \emph{witness distributions}, while a recent definition by
  Sato uses universal quantification over all sets of samples. These notions
  each have their own strengths: the universal version is more general
  than the existential ones, while existential liftings are known to satisfy
  more precise composition principles.  

  We propose a novel, existential version of approximate lifting, called
  $\star$-\emph{lifting}, and show that it is equivalent to Sato's construction
  for discrete probability measures. Our work unifies all known notions of
  approximate lifting, yielding cleaner properties, more general constructions,
  and more precise composition theorems for both styles of lifting, enabling
  richer proofs of differential privacy. We also clarify the relation between
  existing definitions of approximate lifting, and consider more general
  approximate liftings based on $f$-divergences.
\end{abstract}

\maketitle

\section{Introduction}
Differential privacy~\citep{DMNS06} is a strong, rigorous notion of statistical
privacy. Informally, differential privacy guarantees to every individual that
their participation in a database query will have a quantitatively small effect
on the query results, limiting the amount that the query answer depends on their
private data. The definition of differential privacy is parametrized by two
non-negative real numbers, $(\epsilon, \delta)$, which quantify the effect of
individuals on the output of the private query: smaller values give stronger
privacy guarantees. The main strengths of differential privacy lie in its
theoretical elegance, minimal assumptions, and flexibility.

Recently, programming language researchers have developed approaches based on
dynamic analysis, type systems, and program logics for formally proving
differential privacy for programs. (We refer the interested reader to a recent
survey~\citep{BartheGHP16} for an overview of this growing field.)  In this
paper, we consider approaches based on relational program
logics~\citep{BartheKOZ13,BartheO13,OlmedoThesis,BartheFGGHS16,BartheGGHS16,Sato16}.
To capture the quantitative nature of differential privacy, these systems rely
on a quantitative generalization of probabilistic couplings (see, \eg,
\citep{Lindvall02,Thorisson00,Villani08}), called \emph{approximate liftings} or
$(\epsilon, \delta)$-liftings.  Prior works have considered several potential
definitions. While all definitions support compositional reasoning and enable
program logics that can verify complex examples from the privacy literature, the
various notions of approximate liftings have different strengths and weaknesses.

Broadly speaking, the first class of definitions require the existence of one or
two \emph{witness distributions} that ``couple'' the output distributions from
program executions on two related inputs (intuitively, the true database and the
true database omitting one individual's record). The earliest
definition~\citep{BartheKOZ13} supports accuracy-based reasoning for the Laplace
mechanism, while subsequent definitions \citep{BartheO13,OlmedoThesis} support
more precise composition principles from differential privacy and can be
generalized to other notions of distance on distributions. These definitions,
and their associated program logics, were designed for discrete distributions.

In the course of extending these ideas to continuous distributions,
\citet{Sato16} proposes a radically different notion of approximate lifting that
does not rely on witness distributions. Instead, it uses a universal
quantification over all sets of samples. Sato shows that this definition is
strictly more general than the existential versions, but it is unclear (a)
whether the gap can be closed and (b) whether his construction satisfies the
same composition principles enjoyed by some existential definitions.

As a consequence, no single definition is known to satisfy the properties needed
to support all existing formalized proofs of differential privacy. Furthermore,
some of the most involved privacy proofs cannot be formalized at all, as their
proofs require a combination of constructions satisfied by existential or
universal liftings, but not both.

\subsection*{Outline of the paper}

After introducing mathematical preliminaries in Section~\ref{s:bg}, we introduce
our main technical contribution: a new, existential definition of approximate
lifting. This construction, which we call $\star$-\emph{lifting}, is a
generalization of an existing definition by \citet{BartheO13,OlmedoThesis}. The
key idea is to slightly enlarge the domain of witness distributions with a
single generic point, broadening the class of approximate liftings. By a maximum
flow/minimum cut argument, we show that $\star$-liftings are equivalent to
Sato's lifting over discrete distributions. This equivalence can be viewed as an
approximate version of Strassen's theorem~\citep{strassen1965existence}, a
classical result in probability theory characterizing the existence of
probabilistic couplings. We present our definition and the proof of equivalence
in Section~\ref{s:star}.

Then, we show that $\star$-liftings satisfy desirable theoretical properties by
leveraging the equivalence of liftings in two ways. In one direction, Sato's
definition gives simpler proofs of more general properties of $\star$-liftings.
In the other direction, $\star$-liftings---like previously proposed existential
liftings---can smoothly incorporate composition principles from the theory of
differential privacy. In particular, our connection shows that Sato's definition
can use these principles in the discrete case. We describe the key theoretical
properties of $\star$-liftings in Section~\ref{s:props}.

Finally, we provide a thorough comparison of $\star$-lifting with other existing
definitions of approximate lifting in Section~\ref{s:comp}, introduce a
symmetric version of $\star$-lifting that satisfies the so-called advanced
composition theorem from differential privacy~\cite{DRV10} in
Section~\ref{s:sym}, and generalize $\star$-liftings to approximate liftings
based on $f$-divergences in Section~\ref{s:fdiv}.

Overall, the equivalence of $\star$-liftings and Sato's lifting, along with the
natural theoretical properties satisfied by the common notion, suggest that
these definitions are two views on the same concept: an approximate version of
probabilistic coupling.

\section{Background} \label{s:bg}

To model probabilistic data, we work with \emph{discrete sub-distributions}.

\begin{defi}
  A \emph{sub-distribution} over a set $A$ is defined by its
  mass function $\mu : A \to [0, 1]$, which gives the probability of the
  singleton events $a \in A$. This mass function must be s.t.
  $\mass{\mu} \eqdef \sum_{a \in A} \mu(a)$ is well-defined and at most $1$.
  In particular, the \emph{support}
  $\supp(\mu) \eqdef \{ a \in A \mid \mu(a) \neq 0 \}$
  must be discrete (\ie finite or countably infinite).
  When the \emph{weight} $\mass{\mu}$ is equal to $1$, we call $\mu$ a
  \emph{(proper) distribution}.  We let $\Dist(A)$ denote the set of
  sub-distributions over $A$.
  \emph{Events} $E$ are predicates on $A$; the probability of an event $E(x)$
  w.r.t.\ $\mu$, written $\P{x}{\mu}{E(x)}$ or $\sP{\mu}{E}$, is defined as
  $\sum_{x \in A \mid E(x)} \mu(x)$.
\end{defi}

Simple examples of sub-distributions include the \emph{null
  sub-distribution} $\dnull[A] \in \Dist(A)$, which maps each element
of $A$ to $0$, and the \emph{Dirac distribution centered on $x$},
written $\dunit{x}$, which maps $x$ to $1$ and all other elements to
$0$.
One can equip distributions with the usual monadic structure using the Dirac
distributions $\dunit{x}$ for the unit and \emph{distribution
  expectation} $\dlet x \mu {f(x)}$ for the bind; if $\mu$ is a
distribution over $A$ and $f$ has type $A \to \Dist(B)$, then the bind
defines a sub-distribution over $B$ via
$\dlet a \mu {f(a)} : b \mapsto \sum_{a} \mu(a) \cdot f(a)(b)$.

If $f : A \to B$, we can lift $f$ to a function $\dlift f :\Dist(A) \to
\Dist(B)$ as
$\dlift{f}(\mu) \eqdef \dlet a \mu {\dunit{f(a)}}$~;
more explicitly,
$\dlift{f}(\mu) : b \mapsto \P{a}{\mu}{a \in f^{-1}(b)}$.
For instance, when working with sub-distributions over pairs, we have the
probabilistic versions $\dfst : \Dist(A \times B) \to \Dist(A)$ and $\dsnd :
\Dist(A \times B) \to \Dist(B)$ (called the \emph{marginals}) of the usual
projections $\fst$ and $\snd$ by lifting. One can check that the \emph{first}
and \emph{second marginals} $\dfst(\mu)$ and $\dsnd(\mu)$ of a distribution
$\mu$ over $A \times B$ are given by the following equations:
$\dfst (\mu)(a)=\sum_{b\in B} \mu(a,b)$ and
$\dsnd (\mu)(b)=\sum_{a\in A} \mu(a,b)$.
When $f : A \to \Dist(B)$, we will abuse notation and write the lifting $\dlift
f : \Dist(A) \to \Dist(B)$ to mean
$\dlift{f}(\mu) \eqdef \dlet x \mu {f(x)}$;
this is sometimes called the \emph{Kleisli extension} of $f$.

Finally, we will often consider sums of weight functions over sets. If $\alpha :
A \to \RR^{\geq 0}$ maps $A$ to the non-negative real numbers, we write $\alpha[X] \in
\RR^{\geq 0} \cup \{ \infty \}$ for $\sum_{x \in X} \alpha(x)$. Moreover, if $\alpha :
A \times B \to \RR^{\geq 0}$, we write $\alpha[X, Y]$ (resp. $\alpha[x, Y]$, $\alpha[X,
y]$) for $\alpha[X \times Y]$ (resp.  $\alpha[\{ x \} \times Y$, $\alpha[X
\times \{ y \}]$).  Note that for a sub-distribution $\mu \in \Dist(A)$ and an
event $E \subseteq A$, $\sP \mu E = \mu[E]$.

\medskip

We now review the definition of differential privacy.

\begin{defi}[\citet{DMNS06}] 
  Let $\epsilon, \delta \geq 0$ be real parameters. A probabilistic computation
  $M : A \to \Dist(B)$ satisfies $(\epsilon,\delta)$-\emph{differential privacy}
  w.r.t.\ an adjacency relation $\pre \subseteq A \times A$ if for every pair of
  inputs $(a, a') \in \pre$ and every subset of outputs $E \subseteq B$, we have
  \[
    \sP{M(a)}{E} \leq e^\epsilon \cdot
      \sP{M(a')}{E} + \delta .
  \]
\end{defi}

Differential privacy is closely related to a relaxed version of
distance---technically, an $f$-divergence---on distributions.

\begin{defi}[\citet{BartheKOZ13,BartheO13,OlmedoThesis}]
  Let $\epsilon\geq 0$. The $\epsilon$-\emph{DP divergence}
  $\Delta_{\epsilon}(\mu_1,\mu_2)$ between two sub-distributions
  $\mu_1, \mu_2 \in \Dist(B)$ is defined as
  \[\sup_{E \subseteq B}
      \left(\sP{\mu_1}{E} - e^\epsilon \cdot
              \sP{\mu_2}{E}\right) . \]
\end{defi}

Then, differential privacy admits an alternative characterization based on
DP divergence.

\begin{lem} \label{l:dp-epsdist}
  A probabilistic computation $M : A \to \Dist(B)$ satisfies
  $(\epsilon,\delta)$-\emph{differential privacy} w.r.t.\ an
  adjacency relation $\pre \subseteq A \times A$ iff
  $\Delta_{\epsilon} (M(a), M(a')) \leq \delta$ for every pair of
  inputs $(a, a') \in \pre$.
\end{lem}

Our new definition of approximate lifting is inspired by a version of
approximate liftings involving two witness distributions, proposed by
\citet{BartheO13,OlmedoThesis}.

\begin{defi}[\citet{BartheO13,OlmedoThesis}] \label{d:2lift}
  Let $\mu_1 \in \Dist(A)$ and $\mu_2 \in \Dist(B)$ be sub-distributions,
  $\epsilon, \delta \in \RR^{\geq 0}$ and $\rR$ be a binary relation over
  $A$ and $B$. An $(\epsilon, \delta)$-\emph{approximate}
  $2$-\emph{lifting} of $\mu_1$ and $\mu_2$ for $\rR$ is a pair
  $(\mu\lside, \mu\rside)$ of sub-distributions over $A \times B$ s.t.
  \begin{enumerate}
    \item $\dfst(\mu\lside) = \mu_1$ and $\dsnd(\mu\rside) = \mu_2$;
    \item $\Delta_\epsilon(\mu\lside, \mu\rside) \leq \delta$; and
    \item $\supp(\mu\lside) \subseteq \rR$ and $\supp(\mu\rside) \subseteq \rR$.
  \end{enumerate}
  We write
  $
    \mu_1 \alifticalp{\rR}{\epsilon, \delta} \mu_2
  $
  if there exists an $(\epsilon, \delta)$-approximate ($2$-)lifting of $\mu_1
$ and $\mu_2$ for $\rR$; the superscript $\cdot^{(2)}$ indicates that
  there are two witnesses $\mu\lside$ and $\mu\rside$ in this definition of
  lifting.
\end{defi}
Combined with Lemma~\ref{l:dp-epsdist}, a probabilistic computation $M : A \to
\Dist(B)$ is $(\epsilon, \delta)$-differentially private if and only if for
every two adjacent inputs $a \mathrel{\phi} a'$, there is an approximate
lifting of the equality relation: $M(a) \alifticalp{=}{\epsilon, \delta} M(a')$.

$2$-liftings can be generalized by varying the notion of distance given by
$\Delta_\epsilon$; we will return to this point in Section~\ref{s:fdiv}. These
liftings also satisfy useful theoretical properties, but some of the properties
are not as general as we would like. For example, it is known that $2$-liftings
satisfy the following mapping property.

\begin{thm}[\citet{BartheFGGHS16}] \label{t:map:restr}
  Let $\mu_1 \in \Dist(A_1)$, $\mu_2 \in \Dist(A_2)$ be distributions
  $f_1 : A_1 \to B_1$, $f_2 : A_2 \to B_2$ be \emph{surjective} maps and $\rR$
  be a binary relation on $B_1$ and $B_2$. Then
  \[
    \dlift{f_1}(\mu_1) \alifticalp{\rR}{\epsilon,\delta} \dlift{f_2}(\mu_2)
    \iff
    \mu_1 \alifticalp{\rS}{\epsilon,\delta} \mu_2
  \]
  where
  $a_1 \mathrel{\rS} a_2
    \iffdef f_1(a_1) \rR f_2(a_2)$.
\end{thm}

This property can be used to pull back an approximate lifting on two
distributions over $B_1, B_2$ to an approximate lifting on two distributions
over $A_1, A_2$. For applications in program logics, $B_1, B_2$ could be the
domain of a program variable, $A_1, A_2$ could be the set of memories, and $f_1,
f_2$ could project a memory to a program variable. While the mapping theorem is
quite useful, it is puzzling why it only applies to surjective maps. For
instance, this theorem cannot be used when the maps $f_1, f_2$ inject a smaller
space into a larger space.

For another example, there exist $2$-liftings of the following form, sometimes
called the \emph{optimal subset coupling}.

\begin{thm}[\citet{BartheFGGHS16}] \label{t:subset:opt}
  Let $\mu  \in \Dist(A)$ and consider two subsets $P_1 \subseteq P_2 \subseteq
  A$. Suppose that $P_2$ is a \emph{strict subset} of $A$. Then, we have the
  following equivalence:
  \[
    \sP{\mu}{P_2} \leq e^\epsilon \cdot \sP{\mu}{P_1}
    \iff
    \mu \alifticalp{\rR}{\epsilon, 0} \mu ,
  \]
  where
  $a_1 \mathrel{\rR} a_2
    \iffdef a_1 \in P_1 \iff a_2 \in P_2$.
\end{thm}

In this construction, it is puzzling why the larger subset $P_2$ must be a
\emph{strict} subset of the domain $A$. For example, this theorem does not apply
for $P_2 = A$, but we may be able to construct the approximate lifting if we
simply embed $A$ into a larger space $A'$---even though $\mu$ has support over
$A$! Furthermore, it is not clear why the subsets must be nested, nor is it
clear why we can only relate $\mu$ to itself.

These shortcomings suggest that the definition of $2$-liftings may be
problematic. While the distance condition appears to be the most constraining
requirement, the marginal and support conditions are responsible for the main
issues.

\paragraph*{Witnesses must have support in the relation $\rR$.}
For some relations $\rR$, there may be elements $a$ such that $a \mathrel{\rR}
b$ does not hold for any $b$, or vice versa. It can be impossible to find
witnesses with the correct marginals on these elements while satisfying the
support condition, even if the distance condition is satisfied. At a high level,
there are situations where it is possible to construct a pair $\mu\lside$ and
$\mu\rside$ satisfying the distance requirement, but where $\mu\rside$ needs
additional mass to achieve the marginal requirement for some element $b$. Adding
this mass anywhere preserves the distance bound between $\mu\lside$ and
$\mu\rside$---since it only increases the mass of $\mu\rside$ while preserving
the mass of $\mu\lside$, and the distance bound is asymmetric---but if there is
no element $a$ such that $a \mathrel{\rR} b$ then we cannot place this mass
within the required support, and $\mu\lside$ and $\mu\rside$ cannot be related
by an approximate lifting.

\paragraph*{No canonical choice of witnesses.}
A related problem is that the marginal requirement only constrains one marginal
of each witness distribution. Along the other component, the witnesses may place
the mass anywhere on any pair in the relation. As a result, witnesses to an
approximate lifting $\mu_1 \alifticalp{\rR}{\epsilon, \delta} \mu_2$ are
sometimes required to place mass outside of $\supp(\mu_1) \times \supp(\mu_2)$,
even though intuitively only elements in the support of the related
distributions should be relevant to the lifting. This theoretical flaw makes it
difficult to establish basic mapping and support properties of approximate
liftings.

\begin{exa}
  We illustrate these problems with a concrete example. Consider the geometric
  distribution with parameter $p = 1/2$, a distribution over the natural numbers
  which models the distribution of number of flips of a fair coin before the
  coin first comes up tails. Formally, the distribution $\gamma \in \Dist(\NN)$
  is defined by $\gamma(k) = 1 / 2^{k + 1}$. Consider the binary relation $\rR =
  \{ (x_1, x_2) \mid x_1 + 1 = x_2 \}$ over $\NN$. Now, $\gamma$ cannot be
  related to itself via an approximate lifting $\gamma
  \alifticalp{\rR}{\epsilon, \delta} \gamma$ for any parameters $\epsilon,
  \delta$. To see why, the second witness $\mu\rside$ must satisfy the second
  marginal condition at $k = 0$, so it must put total weight $\gamma(0) = 1/2$
  on pairs of the form $(-, 0)$. These pairs must belong to the relation $\rR$,
  but there is no $x_1 \in \NN$ such that $x_1 + 1 = 0$.  However, there
  \emph{is} an approximate lifting $\overline{\gamma}
  \alifticalp{\overline{\rR}}{\ln(2), 0} \overline{\gamma}$, where $\gamma$ and
  $\rR_{+1}$ are extended to the integers $\ZZ$. For instance, the two joint
  distributions with support $\mu\lside(z, z + 1) = 1/2^{z + 1}$ for $z \geq 0$,
  and $\mu\rside(z, z+ 1) = 1/2^{z + 2}$ for $z \geq - 1$ form witnesses.

  This behavior is a sign that the notion of approximate lifting is not
  well-behaved: the support of $\overline{\gamma}$ remains the non-negative
  integers, but somehow embedding $\gamma$ into a larger space enables
  additional approximate liftings.
\end{exa}
\section{%
  \texorpdfstring
  {$\star$-Liftings and Strassen's Theorem}
  {*-Liftings and Strassen's Theorem}}
\label{s:star}

To improve the theoretical properties of $2$-liftings, we propose a simple
extension: allow witnesses to be distributions over a larger set.

\begin{nota}
  For a set $A$, we write $A^\star$ for $A \uplus \{ \star \}$. For a
  distribution $\eta \in \Dist(C)$ and a subset $C' \subseteq C$, we write
  $\eta_{|C'}$ for the \emph{restriction} of $\eta$ to $C'$, i.e., the
  sub-distribution given by $\eta_{|C'}(c) = \eta(c)$ for $c \in C'$, and
  $\eta_{|C'}(c) = 0$ otherwise.
\end{nota}

\begin{defi}[$\star$-lifting]
  Let $\mu_1 \in \Dist(A)$ and $\mu_2 \in \Dist(B)$ be sub-distributions,
  $\epsilon, \delta \in \RR^{\geq 0}$ and $\rR$ be a binary relation over $A$
  and $B$. An $(\epsilon, \delta)$-\emph{approximate} $\star$-\emph{lifting} of
  $\mu_1$ and $\mu_2$ for $\rR$ is a pair of sub-distributions
  $\eta\lside \in \Dist(A \times B^\star)$ and
  $\eta\rside \in \Dist(A^\star \times B)$ s.t.
  \begin{enumerate}
  \item $\dfst(\eta\lside) = \mu_1$ and $\dsnd(\eta\rside) = \mu_2$;
  \item
    $\supp({\eta\lside}_{|A \times B}) ,
     \supp({\eta\rside}_{|A \times B}) \subseteq \oR$; and
  \item $\Delta_{\epsilon}(
           \overline{\eta\lside},
           \overline{\eta\rside}) \leq \delta$,
  where $\overline{\eta_{\bullet}}$ is the extension of
  $\eta_{\bullet}$ to $\Dist(A^\star \times B^\star)$ given by the evident
  inclusions from $A \times B^\star$ and $A^\star \times B$ to $A^\star \times
  B^\star$.
  \end{enumerate}
  We write
  $\mu_1 \aliftnew{R}{\epsilon, \delta} \mu_2$
  if there exists an $(\epsilon, \delta)$-approximate $\star$-lifting of
  $\mu_1$ and $\mu_2$ for $\rR$.
\end{defi}

By adding an element $\star$, we address both problems discussed at
the end of the previous section. First, for every $a \in A$ witnesses may
place mass at $(a, \star)$, and for every $b \in B$ witnesses may place
mass at $(\star, b)$. Second, $\star$ serves as a generic element
where all mass outside the supports $\supp(\mu_1) \times
\supp(\mu_2)$ may located, giving more control over the form of the witnesses.
Formally, $\star$-liftings satisfy the following natural support property.

\begin{lem} \label{l:alift:supp}
  Let $\mu_1 \in \Dist(A)$ and $\mu_2 \in \Dist(B)$ be distributions such that
  $ \mu_1 \aliftnew{\rR}{\epsilon, \delta} \mu_2$ .
  Then, there are witnesses with support contained in $\supp(\mu_1)^\star \times
  \supp(\mu_2)^\star$.
\end{lem}

\proofatend
  Let $\mu\lside$ and $\mu\rside$ be any pair of witnesses to the approximate
  lifting. We will construct witnesses $\eta\lside, \eta\rside$ with the
  desired support. For ease of notation, let $S_i \eqdef \supp(\mu_i)$ for
  $i \in \{ 1, 2 \}$.  Define:
  \[
    \begin{gathered}
    \eta\lside(a, b) =
    \begin{cases}
      \mu\lside(a, b) &: (a, b) \in S_1 \times S_2 \\
      \mu\lside[a, B^\star \setminus S_2] &: b = \star
    \end{cases} \\
    \eta\rside(a, b) =
    \begin{cases}
      \mu\rside(a, b) &: (a, b) \in S_1 \times S_2 \\
      \mu\rside[A^\star \setminus S_1, b] &: a = \star
    \end{cases}
    \end{gathered}
  \]
  Evidently, $\eta\lside$ and $\eta\rside$ have support in $S_1^\star \times
  S_2^\star$. Additionally, it is straightforward to check that
  $\dfst(\eta\lside) = \dfst(\mu\lside) = \mu_1$ and $\dsnd(\eta\rside) =
  \dsnd(\mu\rside) = \mu_2$ so $\eta\lside$ and $\eta\rside$ have the desired
  marginals.

  It only remains to check the distance condition. By the definition of the
  distance $\Delta_\epsilon$, we know that there are non-negative values
  $\delta(a, b)$ such that (i) $\overline{\mu\lside}(a, b) \leq e^\epsilon
  \overline{\mu\rside}(a, b) + \delta(a, b)$ and (ii) $\sum_{a, b} \delta(a, b)
  \leq \delta$. We can define new constants:
  \[
    \zeta(a, b) =
    \begin{cases}
      \delta(a, b) &: (a, b) \in S_1 \times S_2 \cup \{ \star \} \times B \\
      \delta[a, B^\star \setminus S_2] &: b = \star .
    \end{cases}
  \]
  Since $\overline{\mu\lside}(\star, b) = \overline{\eta\lside}(\star, b) = 0$
  for all $b \in B^\star$, and $\overline{\mu\rside}(a, b) =
  \overline{\eta\rside}(a, b) = 0$ for all $b \notin S_2$, point (i) holds for
  the witnesses $\eta\lside, \eta\rside$ and constants $\zeta(a, b)$. Since
  $\sum_{a, b} \zeta(a, b) = \sum_{a, b} \delta(a,b) \leq \delta$, point (ii)
  holds as well. Hence, $\Delta_\epsilon(\overline{\eta\lside},
  \overline{\eta\rside}) \leq \delta$ and we have witnesses for the desired
  approximate lifting.
\endproofatend

\subsection{Basic Properties}

$\star$-liftings satisfy key properties enjoyed by existing notions of
approximate lifting. To start, $\star$-liftings characterize differential
privacy.\footnote{%
  The proofs of the next two lemmas use an equivalence that we will soon prove
in Theorem~\ref{t:alift:dom}. This is purely for convenience---these proofs
could also be performed separately, and in any case Theorem~\ref{t:alift:dom}
does not use these lemmas so there is no circularity.}

\begin{lem} \label{l:alift:dp}
  A randomized algorithm $P : A \to \Dist(B)$ is
  $(\epsilon, \delta)$-differentially private w.r.t.\ $\pre$ if for all
  $(a_1, a_2) \in \pre$ we have
  $P(a_1) \aliftnew{=}{\epsilon,\delta} P(a_2)$.
  (Here, we are turning the equality relation on $B$ to an approximate lifting
  relating distributions over $B$.)
\end{lem}

\proofatend
\begin{description}[style=unboxed,leftmargin=0cm]
 \item[($\Longrightarrow$)]
   Let $P$ be $(\epsilon, \delta)$-differentially private w.r.t.\ $\pre$
   and let $(a_1, a_2) \in \pre$. Let $X$ be a subset of $B$. By
   definition of differential privacy, we have
   $P(a_1)[X] \leq e^\epsilon \cdot P(a_2)[X] + \delta
     = e^\epsilon \cdot P(a_2)[(=)(X)] + \delta$.
   Recall that the image of a set $X$ under a binary relation is simply the set
   of all elements related to some element in $X$. In particular, $(=)(X)$ is
   just $X$. Hence, by application of Theorem~\ref{t:alift:dom}, we have $P(a_1)
   \aliftnew{=}{\epsilon,\delta} P(a_2)$.
 \item[($\Longleftarrow$)]
   By application of Theorem~\ref{t:alift:dom}, we have that
   \[ \forall a_1, a_2 \in A,
        \forall X \subseteq B .\, (a_1,a_2) \in \pre
        \implies P(a_1)[X] \leq e^\epsilon \cdot P(a_2)[X] + \delta
        . \]
   This is the definition of $P$ being
   $(\epsilon, \delta)$-differentially private w.r.t.\ $\pre$. \qedhere
\end{description}
\endproofatend

The next lemma establishes several other basic properties of $\star$-liftings:
monotonicity, and closure under relational and sequential composition.

\begin{lem} \label{l:alift:basics}
  \begin{itemize}
  \item Let $\mu_1 \in \Dist(A)$, $\mu_2 \in \Dist(B)$, and $\rR$
    be a binary relation over $A$ and $B$. If
    $\mu_1 \aliftnew{\rR}{\epsilon,\delta} \mu_2$, then for any
    $\epsilon' \geq \epsilon$, $\delta' \geq \delta$ and
    $\oS \supseteq \oR$, we have
    $\mu_1 \aliftnew{\rS}{\epsilon',\delta'} \mu_2$.

  \item Let $\mu_1 \in \Dist(A)$, $\mu_2 \in \Dist(B)$,
    $\mu_3 \in \Dist(C)$ and $\oR$ (resp. $\oS$) be a binary relation
    over $A$ and $B$ (resp. over $B$ and $C$). If
    $\mu_1 \aliftnew{\rR}{\epsilon,\delta} \mu_2$ and
    $\mu_2 \aliftnew{\rS}{\epsilon',\delta'} \mu_3$,
    then
    $\mu_1 \aliftnew {(\rS \circ \rR)} {\epsilon +
      \epsilon',\delta+ e^\epsilon \cdot\delta'} \mu_3$.

  \item For $i \in \{ 1, 2 \}$, let $\mu_i \in \Dist(A_i)$ and
    $\eta_i : A_i \to \Dist(B_i)$. Let $\rR$ (resp. $\rS$) be a binary
    relation over $A_1$ and $A_2$ (resp. over $B_1$ and $B_2$). If
    $\mu_1 \aliftnew{\rR}{\epsilon,\delta} \mu_2$
    for some $\epsilon, \delta \geq 0$ and for any
    $(a_1, a_2) \mathrel{\in} \rR$, we have
    $\eta_1(a_1) \aliftnew{\rS}{\epsilon',\delta'} \eta_2(a_2)$
    for some $\epsilon', \delta' \geq 0$,
    then
    \[ \sE {\mu_1} {\eta_1}
      \aliftnew{\rS}{\epsilon + \epsilon', \delta + \delta'}
      \sE {\mu_2} {\eta_2} . \]
  \end{itemize}
\end{lem}

\proofatend\strut{}
\begin{itemize}
  \item Immediate.

  \item
  Let
  $\overline{\epsilon} \eqdef \epsilon + \epsilon'$ and
  $\overline{\delta} \eqdef  \delta+ e^\epsilon \cdot\delta'$.
  By Theorem~\ref{t:alift:dom}, it is sufficient to show that
  $\mu_1(X) \leq e^{\overline{\epsilon}} \cdot
     \mu_1(\oS (\oR (X))) + \overline{\delta}$
  for any set $X$. We have:
  \begin{align*}
    \mu_1[X]
      &\leq e^\epsilon \cdot \mu_2[\oR(X)] + \delta 
        \tag{Theorem \ref{t:alift:dom}} \\
      &\leq e^\epsilon \cdot (%
          e^{\epsilon'} \cdot \mu_3[\oS(\oR(X))] + \delta')
        + \delta
        \tag{Theorem \ref{t:alift:dom}} \\
      &= e^{\epsilon + \epsilon'} \cdot \mu_3[\oS(\oR(X))]
         + e^{\epsilon} \cdot \delta' + \delta .
  \end{align*}

  \item
  We know that
  $\exists\ \dalifted {\mu\lside} {\mu\rside}
     \epsilon \delta {\rR} {\mu_1} {\mu_2}$.
  Likewise, for $a \eqdef (a_1, a_2) \in \oR$,
  $\exists\ \dalifted {\eta_{\lmark,a}} {\eta_{\rmark,a}}
     {\epsilon'} {\delta'} {\rS} {\eta_1(a_1)} {\eta_2(a_2)}$.
  Let $\eta\lside$ and $\eta\rside$ be the following distribution
  constructors:
  \begin{align*}
    \eta\lside : a & \mapsto \left\{
      \begin{aligned}
        &\eta_{\lmark,a} & \text{if $a \in \rR$} \\
        &\dnull & \text{otherwise}
      \end{aligned} \right. &
    \eta\rside : a & \mapsto \left\{
      \begin{aligned}
        &\eta_{\rmark,a} & \text{if $a \in \rR$} \\
        &\dnull & \text{otherwise}
      \end{aligned} \right. &
  \end{align*}
  and let
  $\xi\lside \eqdef \sE {\mu\lside} {\eta\lside}$
  (resp. $\xi\rside \eqdef \sE {\mu\rside} {\eta\rside}$).
  We now prove that:
  \[ \dalifted {\xi\lside} {\xi\rside}
     {\epsilon+\epsilon'} {\delta+\delta'} {\rS}
     {\sE {\mu_1} {\eta_1}} {\sE {\mu_2} {\eta_2}} . \]

  The marginal and support conditions are immediate. The distance
  condition is obtained by an immediate application of the previous
  point. \qedhere
\end{itemize}
\endproofatend

\subsection{Equivalence with Sato's Definition}

In recent work on verifying differential privacy over continuous distributions,
\citet{Sato16} proposes an alternative definition of approximate lifting. In the
special case of discrete distributions---where all events are measurable---his
definition can be stated as follows.

\begin{defi}[\citet{Sato16}]
  Let $\mu_1 \in \Dist(A)$ and $\mu_2 \in \Dist(B)$, $\rR$ be a binary relation
  over $A$ and $B$ and $\epsilon, \delta \geq 0$.  The distributions
  $\mu_1$ and $\mu_2$ are related by a \emph{(witness-free)
  $(\epsilon,\delta)$-approximate lifting} for $\rR$ if
  \[
    \forall X \subseteq A .\, \mu_1[X] \leq e^\epsilon \cdot
    \mu_2[\oR(X)] + \delta .
  \]
  We write $\rR(X) = \{ b \in B \mid \exists a \in X.\, (a, b) \in \rR \}
  \subseteq B$.
\end{defi}

Notice that this definition has no witness distributions at all; instead, it
uses a universal quantifier over all subsets.  We can show that
$\star$-liftings are equivalent to Sato's definition in the case of
discrete distributions. This equivalence is reminiscent of Strassen's
theorem from probability theory, which characterizes the existence of
probabilistic couplings.
\begin{thm}[\citet{strassen1965existence}]
  Let $\mu_1 \in \Dist(A)$, $\mu_2 \in \Dist(B)$ be two proper distributions,
  and let $\oR$ be a binary relation over $A$ and $B$. Then there exists
  a joint distribution $\mu \in \Dist(A \times B)$ with support in $\oR$ such
  that $\dfst(\mu) = \mu_1$ and $\dsnd(\mu) = \mu_2$ if and only if
  \[
    \forall X \subseteq A .\, \mu_1[X] \leq \mu_2[\oR(X)] .
  \]
\end{thm}
Our result (Theorem~\ref{t:alift:dom}) can be viewed as a generalization of
Strassen's theorem to approximate couplings. The key ingredient in our proof is
the \emph{max-flow min-cut} theorem; we begin by reviewing the basic setting.

\begin{defi}[Flow network]
  A \emph{flow network} is a structure $((V, E), \ssrc, \sdst, c)$
  s.t.  $\mathcal{N} = (V, E)$ is a loop-free directed graph without
  infinite simple path (or rays), $\ssrc$ and $\sdst$ are two distinct
  distinguished vertices of $\mathcal{N}$ s.t. no edge starts from
  $\sdst$ and ends at $\ssrc$, and
  $c : E \to \mathbb{R}^+ \cup \{ +\infty \}$
  is a function assigning to each edge of $\mathcal{N}$ a capacity. The capacity
  $c$ is extended to $V^2$ by assigning capacity $0$ to any pair
  $(u, v)$ s.t. $(u, v) \notin E$.
\end{defi}

\begin{defi}[Flow]
  Given a flow network $\mathcal{N} \eqdef ((V, E), \ssrc, \sdst, c)$,
  a function $f : V^2 \to \mathbb{R}$ is a \emph{flow} for $\mathcal{N}$
  iff
  \begin{enumerate} \setlength{\itemsep}{0pt}
  \item $\forall u, v \in V .\, f(u, v) \leq c(u, v)$,
  \item $\forall u, v \in V .\, f(u, v) = -f(v, u)$, and
  \item $\forall u \in V .\, u \notin \{ \ssrc, \sdst \} \implies
           \sum_{v \in V} f(u, v) = 0$
        (Kirchhoff's Law).
  \end{enumerate}
  The \emph{mass} $|f|$ of a flow $f$ is defined as
  $|f| \eqdef \sum_{v \in V} f(\ssrc, v) \in \RR \{\cup +\infty\}$.
  %
\end{defi}

\begin{defi}[Cut]
  Given a flow network $\mathcal{N} \eqdef ((V, E), \ssrc, \sdst, c)$,
  a \emph{cut} for $\mathcal{N}$ is any set $C \subseteq V$ that
  partition $V$ s.t. $\ssrc \in V$ but $\sdst \notin V$.  The
  \emph{cut-set} $\cset{C} \subseteq E$ of a cut $C$ is the set of edges
  crossing the cut:
  $\{ (u, v) \in E \mid u \in C, v \notin C \}$.
  The \emph{capacity} $|C| \in \RR^{\geq 0} \cup \{\infty\}$ of a cut $C$ is defined as
  $|C| \eqdef \sum_{(u, v) \in \cset{C}} c(u, v)$.
\end{defi}

For finite flow networks, the maximum flow is equal to the minimum cut (see,
e.g., \citet{Kleinberg:2005:AD:1051910}).
\citet{DBLP:journals/jct/AharoniBGPS11} generalize this theorem to networks with
countable vertices and edges under certain conditions. We will use a consequence
of their result.

\begin{thm}[Weak Countable Max-Flow Min-Cut]
  \label{t:maxflow}
  Let $\mathcal{N}$ be flow network with (a) no infinite directed paths and (b)
  finite total capacity leaving $\top$ and entering $\bot$. Then,
  \[ \sup \{ |f| \mid \text{$f$ is a flow for $\mathcal{N}$} \} =
     \inf \{ |C| \mid \text{$C$ is a cut for $\mathcal{N}$} \} \]
  and both supremum and infimum are achieved by some flow and cut, respectively.
\end{thm}

We are now ready to prove an approximate version of Strassen's theorem, thereby
showing equivalence between $\star$-liftings and Sato's liftings.

\begin{thm} \label{t:alift:dom}
  Let $\mu_1 \in \Dist(A)$ and $\mu_2 \in \Dist(B)$, $\rR$ be a binary
  relation over $A$ and $B$ and $\epsilon, \delta \in \RR^{\geq 0}$.  Then,
  $\mu_1 \aliftnew{R}{\epsilon, \delta} \mu_2$
  iff
  $\forall X \subseteq A .\, \mu_1(X) \leq e^\epsilon \cdot
  \mu_2(\oR(X)) + \delta$.
\end{thm}

\begin{figure}
  \centering
  \begin{tikzpicture}[fontscale=-1,>={latex}]
  \tikzstyle{node_style}=[draw,circle,minimum size=1.5em,inner sep=1]
  \tikzset{snake it/.style={decorate, decoration=snake}}
  \useasboundingbox (-0.5,0.5) rectangle (4,-3.3);
  \begin{scope}[transform canvas={scale=0.9}]
    \begin{scope}[auto, every node/.style={node_style},node distance=0.4 cm]
    \node (as) at (0, 0) {$\star^{\ssrc}$};
    \node[below=0.3cm of as] (a1) {$a_1^{\ssrc}$};
    \node[below=0.3cm of a1] (a2) {$a_2^{\ssrc}$};
    \node[below=0.6cm of a2] (an) {$a_n^{\ssrc}$};

    \node[right=5.0cm of as] (bs) {$\star^{\sdst}$};
    \node[below=0.3cm of bs] (b1) {$b_1^{\sdst}$};
    \node[below=0.3cm of b1] (b2) {$b_2^{\sdst}$};
    \node[below=0.6cm of b2] (bn) {$b_n^{\sdst}$};

    \coordinate[between=a1 and an] (mida);
    \coordinate[between=b1 and bn] (midb);

    \node[left =1.5cm of mida] (src) {$\ssrc$};
    \node[right=1.5cm of midb] (dst) {$\sdst$};

    \coordinate[between=a1 and an] (smida);
    \coordinate[between=b1 and bn] (smidb);

    \coordinate[between=smida and smidb] (mid);
    \end{scope}

    \node
      [cloud, draw, cloud puffs=20, minimum height=3cm, align=center]
      at (mid) (R)
      {$\begin{array}{@{}c@{}}
          a_i \rR b_j \\
          \Downarrow  \\
          \infty
        \end{array}$};

    \path[->] (src) edge node[above] {} (as);
    \path[->] (src) edge node[above] {} (a1);
    \path[->] (src) edge node[above] {} (a2);
    \path[->] (src) edge node[above] {} (an);

    \path[-] (a1) edge node {} (R);
    \path[-] (a2) edge node {} (R);
    \path[-] (an) edge node {} (R);

    \path[->] (as) edge[bend left=20]
      node[bend left=20, above] {$\infty$} (bs);

    \draw[->] (as) to[bend left=30] (b1);
    \draw[->] (as) to[bend left=35] (b2);
    \draw[->] (as) to[bend left=40] (bn);

    \draw[->] (a1) to[bend left=30] (bs);
    \draw[->] (a2) to[bend left=35] (bs);
    \draw[->] (an) to[bend left=40] (bs);

    \path[->] (R) edge node {} (b1);
    \path[->] (R) edge node {} (b2);
    \path[->] (R) edge node {} (bn);

    \path[->] (bs) edge node [above] {} (dst);
    \path[->] (b1) edge node [above] {} (dst);
    \path[->] (b2) edge node [above] {} (dst);
    \path[->] (bn) edge node [above] {} (dst);

    \draw[dotted] (a2) -- (an);
    \draw[dotted] (b2) -- (bn);

    \draw[dotted] (an) -- +(0,-0.7);
    \draw[dotted] (bn) -- +(0,-0.7);

    \coordinate (p0) at (src |- an);
    \coordinate (pd) at (src |- as);
    \coordinate[between=src and as] (ps);
    \coordinate[between=src and a1] (p1);
    \coordinate[between=src and a2] (p2);
    \coordinate[between=src and an] (pn);

    \draw[black!30,->] (p0) to[bend right=10] (p1);
    \draw[black!30,->] (p0) to[bend right=12] (p2);
    \draw[black!30,->] (p0) to[bend right=16] (pn);

    \node[below = 0cm of p0] {$e^{-\epsilon} \cdot \mu_1(a_i)$};

    \draw[black!30,->] (pd) to[bend left=16] (ps);
    \node[above = 0cm of pd] {$\omega - e^{-\epsilon} \mass{\mu_1}$};

    \coordinate (q0) at (dst |- bn);
    \coordinate (qd) at (dst |- bs);
    \coordinate[between=dst and bs] (qs);
    \coordinate[between=dst and b1] (q1);
    \coordinate[between=dst and b2] (q2);
    \coordinate[between=dst and bn] (qn);

    \draw[black!30,->] (q0) to[bend left=10] (q1);
    \draw[black!30,->] (q0) to[bend left=12] (q2);
    \draw[black!30,->] (q0) to[bend left=16] (qn);

    \node[below = 0cm of q0] {$\mu_2(b_j)$};

    \draw[black!30,->] (qd) to[bend right=16] (qs);
    \node[above = 0cm of qd] {$e^{-\epsilon} \delta$};
  \end{scope}
  \end{tikzpicture}
  \caption{\label{f:cntnetdiag} Flow Network in Theorem~\ref{t:alift:dom}}
\end{figure}

\begin{proof}
  We detail the reverse direction; the forward direction is immediate.  We can
  assume that $A$ and
  $B$ are countable; in the case where $A$ and $B$ are not both
  countable, we first consider the restriction of $\mu_1$ and $\mu_2$
  to their respective supports---which are countable sets---and
  construct witnesses to the $\star$-lifting. The witnesses can then
  be extended to an approximate coupling of $\mu_1$ and $\mu_2$ by adding a null
  mass to the extra points.

  Let $\omega \eqdef \mass{\mu_2} + e^{-\epsilon}\cdot\delta$ and let
  $\ssrc$ and $\sdst$ be fresh symbols. For any set $X$, define
  $X^{\ssrc}$ and $X^{\sdst}$ resp. as $\{ x^{\ssrc} \mid x \in X \}$
  and $\{ x^{\sdst} \mid x \in X \}$.
  Let $\mathcal{N}$ be the flow network of Figure~\ref{f:cntnetdiag}
  whose resp. source and sink are $\ssrc$ and $\sdst$, whose set of
  vertices $V$ is
  $\{ \ssrc, \sdst \} \uplus (A^\star)^{\ssrc} \uplus (B^\star)^{\sdst}$,
  and whose set of edges $E$ is
  $E_{\ssrc} \uplus E_{\sdst} \uplus E_{\oR} \uplus E_{\star}$ with
  \begin{align*}
    E_{\ssrc} &\eqdef \{ \ssrc \mapsto_{e^{-\epsilon}\mu_1(a)} a^{\ssrc} \mid a \in A \}&
    E_{\sdst} &\eqdef \{ b^{\sdst} \mapsto_{\mu_2(b)} \sdst \mid b \in B \}\\
    E_{\oR} &\eqdef \{ a^{\ssrc}  \mapsto_{\infty} b^{\sdst} \mid a \rR b  \vee a = \star \vee b = \star \}&
    E_{\star} &\eqdef \{\ssrc \mapsto_{(\omega - e^{-\epsilon}\mass{\mu_1})} \star^{\ssrc},~\star^{\sdst} \mapsto_{e^{-\epsilon}\delta} \sdst \}.
  \end{align*}
  Let $C$ be a cut of $\mathcal{N}$. In the following, we sometimes use $C$
  to denote both the cut $C$ and its cut-set $\cset{C}$. We check
  $\mass{C} \geq \omega$.  If $C \cap E_{\oR} \neq \emptyset$ then
  $\mass{C} = \infty$.
  Note that $C \cap E_{\star} = \emptyset$ implies
  $C \cap E_{\oR} \neq \emptyset$.
  If $(\ssrc, \star^{\ssrc}) \in C$ and
  $(\sdst, \star^{\sdst}) \notin C$ then we must have
  $E_{\ssrc} \subseteq C$.  This implies that $\mass{C}\geq \omega$
  since $E_{\ssrc} \uplus \{(\ssrc, \star^{\ssrc})\}$ is a cut with
  capacity $\omega$.  If $(\ssrc, \star^{\ssrc}) \notin C$ and
  $(\sdst, \star^{\sdst}) \in C$ then we have
  $\mass{C}\geq \omega$ in the similar way as above.
  Otherwise (\ie $C \cap E_{\oR} = \emptyset$ and
  $E_{\star} \subseteq C$), for $C$ to be a cut, we must have
  $\oR(A \setminus A^{\dagger}) \subseteq B^{\dagger}$ where
  $A^\dagger \eqdef \{ x \in A \mid (\ssrc, x^{\ssrc}) \in C \}$
  and
  $B^\dagger \eqdef \{ y \in B \mid (y^{\sdst}, \sdst) \in C \}$.
  Thus,
  \begin{align*}
   \mass{C}
    &= e^{-\epsilon}\cdot\mu_1[A^\dagger] + \mu_2[B^\dagger] + \mass{E_{\star}} \\
    &\geq e^{-\epsilon}\cdot\mu_1[A^\dagger]
        + \mu_2[\oR(A \setminus A^{\dagger})]
        + e^{-\epsilon}\cdot \delta
        + (\omega - e^{-\epsilon}\cdot\mass{\mu_1}) \\
    &\geq e^{-\epsilon}\cdot(\mu_1[A^\dagger]
        + \mu_1[A \setminus A^{\dagger}])
        + \omega - e^{-\epsilon}\cdot\mass{\mu_1}
     = \omega.
  \end{align*}
  Hence, $E_{\ssrc} \uplus \{(\star^{\sdst}, \sdst)\}$ is a minimum
  cut with capacity $\omega$.
  By Theorem~\ref{t:maxflow},
  we obtain a maximum flow $f$ with mass $\omega$.
  Note that the flow $f$ saturates the capacity of
  all edges in $E_{\ssrc}$, $E_{\sdst}$, and $E_{\star}$.
  Let $\hat{f} : (a, b) \in A^\star \times B^\star \mapsto f(a^{\ssrc}, b^{\sdst})$.
  We now define the following distributions:
  \[ \begin{array}{ll}
  \begin{aligned}
    \eta\lside : A \times B^\star &\to \RR^{\geq 0} \\
      (a, b) &\mapsto e^{\epsilon} \cdot \hat{f}(a, b)
  \end{aligned} &
  \begin{aligned}
    \eta\rside : A^\star \times B &\to \RR^{\geq 0} \\
      (a, b) &\mapsto \hat{f}(a, b).
  \end{aligned}
  \end{array}\]
  We clearly have $\dfst(\eta\lside) = \mu_1$ and
  $\dsnd(\eta\rside) = \mu_2$.
  Moreover, by construction of the
  flow network $\mathcal{N}$, $\supp(\hat{f}_{|A \times B}) \subseteq \oR$. Hence,
  $\supp({\eta\lside}_{|A \times B}) ,
   \supp({\eta\rside}_{|A \times B}) \subseteq \oR$.
  It remains to show that
  $\Delta_{\epsilon}(\overline{\eta\lside}, \overline{\eta\rside}) \leq \delta$.
  Let $X$ be a subset of $A^\star \times B^\star$.
  Let
  $\overline{X_a} \eqdef \{ a \in A \mid (a, \star) \in X \}$,
  $\overline{X_b} \eqdef \{ b \in B \mid (\star, b) \in X \}$ and
  $\overline{X} \eqdef X \cap (A \times B)$.
  Then,
  \begingroup
  \allowdisplaybreaks
  \begin{align*}
  \overline{\eta\lside}[X] - e^{\epsilon} \cdot \overline{\eta\rside}[X]
  &= e^{\epsilon} \left(\hat{f}[\overline{X}] 
     + \hat{f}[\overline{X_a} \times \{\star\}]\right)
     - e^{\epsilon} \left(\hat{f}[\overline{X}] 
     + \hat{f}[\{\star\} \times \overline{X_b}]\right)\\
  &\leq e^{\epsilon} \cdot \hat{f}[\overline{X_a} \times \{\star\}]
  \leq e^{\epsilon} \cdot \hat{f}[A \times \{\star\}]
   = \delta.
  \end{align*}
  \endgroup
  The last equality holds by Kirchhoff's law:
  $\hat{f}[A \times \{\star\}]
   = \sum_{a \in A} f({a^{\ssrc}},{\star^{\sdst}})
   = f({\star^{\sdst}}, \sdst)
   = e^{-\epsilon}\cdot\delta$.
\end{proof}

\xproofatend
  To show the forward direction of Theorem~\ref{t:alift:dom}, let $X \subseteq A$
  and $\rcomp{\oR(X)} = B \setminus \oR(X)$.
  Then, we have
  \begin{align*}
    \mu_1[X]
      &= \dfst(\eta\lside)[X]
       = \eta\lside[X, B^\star]
       = \overline{\eta\lside}[X, B^\star]
       = \overline{\eta\lside}
           [X, \oR(X) \uplus \rcomp{\oR(X)} \uplus \{ \star \}] \\
      &= \overline{\eta\lside}[X, \oR(X) \uplus \{ \star \}]
           + \underbrace{\overline{\eta\lside}[X, \rcomp{\oR(X)}]}_{=\ 0}
       \leq e^{\epsilon} \cdot \overline{\eta\rside}
          [X, \oR(X) \uplus \{ \star \}] + \delta \\
      &\leq
          e^{\epsilon} \cdot \underbrace{%
            \overline{\eta\rside}[A^\star, \oR(X)]}_{=\ \eta\rside[A^\star, \oR(X)]} +
          e^{\epsilon} \cdot \underbrace{%
            \overline{\eta\rside}[A^\star, \{ \star \}]}_{=\ 0} + \delta \\
      &= e^{\epsilon} \cdot \dsnd(\eta\rside)[\oR(X)] + \delta
       = e^{\epsilon} \cdot \mu_2[\oR(X)] + \delta ,
  \end{align*}
  as desired.
\xendproofatend

\subsection{Alternative Proof of Approximate Strassen's Theorem}

We can provide an alternative, arguably simpler proof of the reverse direction
of the approximate Strassen's theorem (Theorem~\ref{t:alift:dom}). Instead of
relying on the max-flow min-cut theorem for countable networks by
\citet{DBLP:journals/jct/AharoniBGPS11}, we apply the more standard result on
finite networks and then pass from approximate liftings on finite restrictions
of the two target distributions to an approximate lifting of the limit
distributions, via a limiting argument. The results of this section have been
formalized in the \textsc{Coq} proof
assistant.\footnote{\url{https://github.com/strub/xhl}}

We first start with a simple technical lemma.

\begin{lem} \label{l:alift:extend}
  Let $\mu_1 \in \Dist(A)$ and $\mu_2 \in \Dist(B)$ (with respective
  support $S_1$ and $S_2$), $\rR$ be a binary relation over
  $A$ and $B$ and $\epsilon, \delta \in \RR^{\geq 0}$.  Then,
  $\drestr{(\mu_1)}{S_1} \aliftnew{R}{\epsilon, \delta} \drestr{(\mu_2)}{S_2}$
  implies
  $\mu_1 \aliftnew{R}{\epsilon, \delta} \mu_2$.
\end{lem}

\proofatend
  If $\eta_L \in \Dist(S_1 \times S_2^\star)$,
  $\eta_R \in \Dist(S_1^\star \times S_2)$ are the witnesses of
  $\drestr{(\mu_1)}{S_1} \aliftnew{R}{\epsilon, \delta} \drestr{(\mu_2)}{S_2}$,
  their extensions to $\Dist(A \times B^\star)$ and
  $\Dist(A^\star \times B)$, padding with $0$, form witnesses for
  $\mu_1 \aliftnew{R}{\epsilon, \delta} \mu_2$.
\endproofatend

We can now prove the theorem for distributions over finite domains:

\begin{lem}[Finite approximate Strassen's theorem]
  \label{l:strassen:fin}
  Let $\mu_1 \in \Dist(A)$ and $\mu_2 \in \Dist(B)$ be sub-distributions with
  finite supports, $\rR$ be a binary relation over $A$ and $B$ and
  $\epsilon, \delta \in \RR^{\geq 0}$. If for all $X \subseteq A$ we have $\mu_1[X]
  \leq e^\epsilon \cdot \mu_2[\oR(X)] + \delta$, then $\mu_1
  \aliftnew{R}{\epsilon, \delta} \mu_2$.
\end{lem}

\proofatend
  The case for distributions with finite domains can be proven using
  the same proof of Theorem~\ref{t:alift:dom}, using the standard
  (finite) max-flow min-cut theorem. The result extends to
  distributions with finite supports via Lemma~\ref{l:alift:extend}.
\endproofatend

To extend the finite case to distributions over countable sets, we need a lemma
that will allow us to assume that the witnesses are within a multiplicative
factor of each other, except possibly on pairs with $\star$.

\begin{lem} \label{l:witdom}
  Let $\mu_1 \in \Dist(A)$ and $\mu_2 \in \Dist(B)$ s.t.
  $\mu_1 \aliftnew{\rR}{\epsilon, \delta} \mu_2$.
  Then, there exists $\eta\lside$ and $\eta\rside$ witnessing the
  lifting s.t. for $(a, b) \in A \times B$, we have:
  \[ \eta\rside(a, b)
       \leq \eta\lside(a, b)
       \leq e^\epsilon \cdot \eta\rside(a, b) . \]
\end{lem}

\proofatend
  Let $\mu\lside$, $\mu\rside$ be witnesses of
  $\mu_1 \aliftnew{\rR}{\epsilon, \delta} \mu_2$.
  For $(a, b) \in A \times B$, let
  \begin{align*}
    \nu\lside(a, b) &\eqdef \min(\mu\lside(a, b), e^\epsilon \cdot \mu\rside(a, b))\\
    \nu\rside(a, b) &\eqdef \min(\mu\lside(a, b), \mu\rside(a, b))
  \end{align*}
  and define $\eta\lside$, $\eta\rside$ as
  \begin{align*}
    \eta\lside &: (a, b) \in A \times B^\star \mapsto \begin{cases}
      \nu\lside(a, b) & \text{if $b \neq \star$}\\
      \mu_1(a) - \sum_{x \in B} \nu\lside(a, x) & \text{otherwise,}
    \end{cases} \\
    \eta\rside &: (a, b) \in A^\star \times B \mapsto \begin{cases}
      \nu\rside(a, b) & \text{if $a \neq \star$}\\
      \mu_2(b) - \sum_{x \in A} \nu\rside(x, b) & \text{otherwise}.
    \end{cases}
  \end{align*}

  Note that $\eta\lside$ and $\eta\rside$ are well-defined as
  sub-distributions by the marginal conditions for $\mu\lside$ and
  $\mu\rside$. To show that the witnesses are non-negative, let $a \in A$ and $b
  \in B$. We have:
  \begin{align*}
    \eta\lside(a, \star)
      &=    \mu_1(a) - \sum_{b \in B} \nu\lside(a, b)
       \geq \mu_1(a) - \sum_{b \in B} \mu\lside(a, b) \\
      &\geq \mu_1(a) - \sum_{b \in B^\star} \mu\lside(a, b)
       =    \mu_1(a) - \mu_1(a) = 0, \\
    \eta\rside(\star, b)
      &=    \mu_2(b) - \sum_{a \in A} \nu\rside(a, b)
       \geq \mu_2(b) - \sum_{a \in A} \nu\rside(a, b) \\
      &\geq \mu_2(b) - \sum_{a \in A^\star} \mu\rside(a, b)
       =    \mu_2(b) - \mu_2(b) = 0.
  \end{align*}
  To show that the witnesses sum to at most $1$, we have
  \begin{align*}
    \sum_{(a, b) \in A \times B^\star} \eta\lside(a, b)
      &= \sum_{a \in A} \mu_1(a) = |\mu_1| \leq 1 , \\
    \sum_{(a, b) \in A^\star \times B} \eta\rside(a, b)
      &= \sum_{b \in B} \mu_2(b) = |\mu_2| \leq 1 .
  \end{align*}
  Moreover, these witness distributions satisfy the claimed distance condition.
  Indeed for $(a, b) \in A \times B$, we have:
  \begin{subequations} \label{eq:ordw}
    \begin{alignat}{1}
    \eta\rside(a, b)
      &=    \min(\mu_1(a), \mu_2(b))
       \leq \min(\mu_1(a), e^\epsilon \cdot \mu_2(b))
       =    \eta\lside(a, b) \\
    \eta\lside(a, b)
      &=    \min(\mu_1(a), e^\epsilon \cdot \mu_2(b))
       \leq \min(e^\epsilon \cdot \mu_1(a), e^\epsilon \cdot \mu_2(b))
       =    e^\epsilon \cdot \eta\rside(a, b) .
    \end{alignat}
  \end{subequations}

  It remains to prove that $\eta\lside$, $\eta\rside$ are witnesses
  for $\mu_1 \aliftnew{\rR}{\epsilon, \delta} \mu_2$. The marginals
  conditions are obvious. For the support condition, let $a \in A$ and
  $b \in B$. Then, $\eta\lside(a, b) > 0$ (resp.
  $\eta\rside(a, b) > 0$) implies $\mu\lside(a, b) > 0$ (resp.
  $\eta\rside(a, b) > 0$) and hence that $a \rR b$.

  The distance condition follows by a calculation.  For $X \subseteq A^\star
  \times B^\star$, we have:
  \begin{align*}
    \overline{\eta\lside}[X]
      &= \eta\lside[X \cap (A \times B)]
           + \eta\lside[X \cap (A \times \{ \star \})] \\
      &\leq e^\epsilon \cdot \eta\rside[X \cap (A \times B)]
           + \eta\lside[X \cap (A \times \{ \star \})]
         \tag{By Eq.~\eqref{eq:ordw}} \\
      &\leq e^\epsilon \cdot \overline{\eta\rside[X]}
           + \eta\lside[X \cap (A \times \{ \star \})] .
  \end{align*}
  To conclude, it suffices to show that
  $\eta\lside[X \cap (A \times \{ \star \})] \leq \delta$.
  For that, let
  \begin{align*}
    \zeta(a, b)
      &\eqdef \max(\overline{\mu\lside}(a, b)
         - e^\epsilon \cdot \overline{\mu\rside}(a, b), 0) .
  \end{align*}
  Then, we can bound
  \begin{align*}
    \eta\lside & [X \cap (A \times \{ \star \})]
       = \sum_{a \in A} \eta\lside(a, \star)
       = \sum_{a \in A} \left( \mu_1(a) - \sum_{b \in B} \nu\lside(a, b) \right) \\
      &= \sum_{a \in A} \left( \mu_1(a) - \sum_{b \in B} \mu\lside(a, b) - \zeta(a, b) \right) \\
      &= \sum_{a \in A} \left( \mu_1(a) - \sum_{b \in B} \mu\lside(a, b) \right)
           + \sum_{(a, b) \in A \times B} \zeta(a, b) \\
      &= \sum_{a \in A} \left( \sum_{b \in B^\star} \mu\lside(a, b)
               - \sum_{b \in B} \mu\lside(a, b)
           \right) + \sum_{(a, b) \in A \times B} \zeta(a, b) \\
      &= \sum_{a \in A} \underbrace{\, \mu\lside(a, \star) \,}_{= \zeta(a, \star)}
           + \sum_{(a, b) \in A \times B} \zeta(a, b)
       = \sum_{(a, b) \in A \times B^\star} \zeta(a, b) \\
      &\leq \sum_{(a, b) \in A ^\star \times B^\star} \zeta(a, b) .
  \end{align*}

  Now, let
  $S \eqdef \{ (a, b) \in A^\star \times B^\star \mid e^\epsilon \cdot
  \mu\rside(a, b) < \mu\lside(a, b) \}$.
  By the distance condition on $\mu\lside$ and $\mu\rside$, we have
  $\overline{\mu\lside}[S] - e^\epsilon \cdot \overline{\mu\rside}[S] \leq \delta$.
  Hence we conclude the desired distance condition:
  \begin{align*}
    \sum_{(a, b) \in A ^\star \times B^\star} \zeta(a, b)
      &= \sum_{(a, b) \in S} \zeta(a, b) + \sum_{(a, b) \notin S} \underbrace{
           \; \zeta(a b) \;}_{=\ 0} \\
      &= \sum_{(a, b) \in S} \overline{\mu\lside}(a, b) -
           e^\epsilon \cdot \overline{\mu\rside}(a, b) \\
      &= \overline{\mu\lside}[S] - e^\epsilon \cdot \overline{\mu\rside}[S]
       \leq \delta . \qedhere
  \end{align*}
\endproofatend

We can now prove the reverse direction of Theorem~\ref{t:alift:dom}.

\begin{proof}[Alternative proof of Theorem~\ref{t:alift:dom}]
  By Lemma~\ref{l:alift:extend}, without loss of generality, we can
  assume that $A$ and $B$ are countable. Hence, there exists a family
  $\{ A_n \}_n$ (resp.\ $\{ B_n \}_n$) of increasing finite subsets of
  $A$ s.t. $\cup_i A_i = A$ (resp.\ of $B$ s.t.  $\cup_i B_i = B$). For
  $n \in \NN$, we denote by $\mu_1^n$ and $\mu_2^n$ the domain
  restrictions of $\mu_1$ to $A_n$ and $\mu_2$ to $B_n$,
  i.e., $\mu_1^n(a) = \mu_1(a)$ if $a \in A_n$, $0$ otherwise and
  $\mu_2^n(b) = \mu_2(b)$ if $b \in B_n$, $0$ otherwise.

  \medskip

  Fix $n \in \NN$ and let $X \subseteq A$. We have:
  \begin{align*}
    \mu_1^n[X]
      &\leq \mu_1[X] \leq e^\epsilon \cdot \mu_2[\oR(x)] + \delta \\
      &= e^\epsilon \cdot (\mu_2[\oR(X) \cap B_n]
           + \mu_2[\oR(X) \cap \rcomp{B_n}]) + \delta \\
      &= e^\epsilon \cdot \mu_2^n[\oR(X)] + \underbrace{
           (e^\epsilon \cdot \mu_2[\oR(X) \cap \rcomp{B_n}] + \delta)%
         }_{\eqdef\ \delta_n}
  \end{align*}

  Hence, by Lemma~\ref{l:strassen:fin}, we have
  $\mu_1^n \aliftnew{\rR}{\epsilon, \delta_n} \mu_2^n$. By
  Lemma~\ref{l:witdom}, we can moreover assume that
  $\mu_1^n \aliftnew{\rR}{\epsilon, \delta_n} \mu_2^n$ is witnessed by
  sub-distributions $\eta\lside^n$ and $\eta\rside^n$ such that
  \begin{gather}
    \forall a \in A, b \in B .\,
      \eta\rside^n(a, b) \leq \eta\lside(a, b)
        \leq e^\epsilon \cdot \eta\rside(a, b) .
    \label{eq:lwitord}
  \end{gather}
  Since $\rR(X) \cap \rcomp{B_n} \xrightarrow[n \to \infty]{} \emptyset$, we
  also have $\delta_n \xrightarrow[n \to \infty]{} \delta$.

  As a countable product of a sequentially compact sets,
  $[0,1]^{A \times B^\star}$ and $[0,1]^{A^\star \times B}$ are
  sequentially compact, and we can find a subsequence of indices
  $\{ \omega_n \}_{n \in \NN}$ s.t. both $\eta\lside^{\omega_n}$,
  $\eta\rside^{\omega_n}$ resp. converge pointwise to
  sub-distributions $\eta\lside$ and $\eta\rside$.
  
  \medskip

  We now prove that these sub-distributions are witnesses for
  $\mu_1 \aliftnew{\rR}{\epsilon, \delta} \mu_2$. It is clear that the
  supports of $\drestr{\eta\lside}{A \times B}$ and
  $\drestr{\eta\rside}{A \times B}$ are included in $\rR$. We now
  detail the marginals and distance conditions.
  First, note that
  \begin{alignat*}{2}
    \eta\lside^n(a,b)
      &\leq \sum_{b \in B^\star} \eta\lside^n(a,b)
       = \dfst(\eta\lside^n)(a) = \mu_1^n(a) \leq \mu_1(a)
      & \quad [a \in A, b \in B^\star]\\
    \eta\rside^n(a,b)
      &\leq \sum_{a \in A^\star} \eta\rside^n(a,b)
       = \dsnd(\eta\rside^n)(b) = \mu_2^n(b) \leq \mu_2(b)
      & \quad [a \in A^\star, b \in B] .
  \end{alignat*}
  Hence, Equation~\eqref{eq:lwitord} yields
  \begin{subequations}
    \begin{alignat}{2}
    \eta\lside^n(a, b) \label{eq:lord1}
      &\leq e^\epsilon \cdot \eta\rside^n(a, b)
       \leq e^\epsilon \cdot \mu_2(b)
      & \quad [a \in A, b \in B^\star] \\
    \eta\rside^n(a, b) \label{eq:lord2}
      & \leq \eta\lside^n(a, n)
        \leq \mu_1(a)
      & \quad [a \in A^\star, b \in B] .
    \end{alignat}
  \end{subequations}

  \bigskip

  Now, for the first marginal, let $a \in A$. We have:
  \begin{align*}
    \dfst(\eta\lside)(a)
      &= \sum_{b \in B^\star} \eta\lside(a, b)
       = \sum_{b \in B^\star} \lim_{n \to \infty} \eta\lside^{\omega_n}(a, b)
  \end{align*}
  From \eqref{eq:lord1}, it is clear that
  $b \in B^\star \mapsto \eta\lside^{\omega_n}(a, b)$ is absolutely
  dominated by the summable function
  $[b \in B^\star \mapsto \text{$e^\epsilon \cdot \mu_2(b)$ if $b \in B$, else $1$}]$.
  By the dominated convergence theorem, we can swap the limit and summation:
  \begin{align*}
    \dfst(\eta\lside)(a)
      &= \lim_{n \to \infty} \sum_{b \in B^\star} \eta\lside^{\omega_n}(a,b)
       = \lim_{n \to \infty} \dfst(\eta\lside)(a) \\
     & = \lim_{n \to \infty} \mu_1^{\omega_n}(a) = \mu_1(a) .
  \end{align*}

  \medskip

  For the second marginal, let $b \in B$. We have:
  \begin{align*}
    \dsnd(\eta\rside)(b)
      &= \sum_{a \in A^\star} \eta\rside(a, b)
       = \sum_{a \in A^\star} \lim_{n \to \infty} \eta\rside^{\omega_n}(a, b)
  \end{align*}
  From \eqref{eq:lord2}, we have that
  $a \in A^\star \mapsto \eta\lside^{\omega_n}(a, b)$ is absolutely
  dominated by the summable function:
  $[a \in A^\star \mapsto \text{$\mu_1(a)$ if $a \in A$, else $1$}]$.
  Again by the dominated convergence theorem, we can swap the limit and
  summation:
  \begin{align*}
    \dsnd(\eta\rside)(b)
      &= \lim_{n \to \infty} \sum_{a \in A^\star} \eta\rside^{\omega_n}(a,b)
       = \lim_{n \to \infty} \dsnd(\eta\rside)(b) \\
     & = \lim_{n \to \infty} \mu_2^{\omega_n}(b) = \mu_1(b) .
  \end{align*}

  We are left to prove the distance condition, i.e., that for any
  $X \subseteq A^* \times B^*$, we have
  $ \overline{\eta\lside}[X]
       - e^\epsilon \cdot \overline{\eta\rside}[X] \leq \delta $.
  First, note that
  $\overline{\eta\lside}[X]
     = \lim_{n \to \infty} \overline{\eta\lside}^{\omega_n}[X]$
  and
  $\overline{\eta\rside}[X]
     = \lim_{n \to \infty} \overline{\eta\rside}^{\omega_n}[X]$.
  Indeed, for $\overline{\eta\lside}[X]$, we have
  \begin{align*}
    \overline{\eta\lside}[X]
      &= \sum_{a, b \in X} \lim_{n \to \infty} \eta\lside^{\omega_n}(a, b)
       = \sum_{a \in A^\star} \sum_{b \in X\lside^a} \lim_{n \to \infty} \eta\lside^{\omega_n}(a, b) \\
      &= \sum_{a \in A^\star} \lim_{n \to \infty} \sum_{b \in X\lside^a} \eta\lside^{\omega_n}(a, b)
           \tag{dominated convergence} \\
      &= \lim_{n \to \infty} \sum_{a \in A^\star} \sum_{b \in X\lside^a} \eta\lside^{\omega_n}(a, b)
           \tag{dominated convergence} \\
      &= \lim_{n \to \infty} \sum_{a, b \in X} \eta\lside^{\omega_n}(a, b)
       = \lim_{n \to \infty} \eta\lside^{\omega_n}(a, b)[X]
  \end{align*}
  where $X\lside^a \eqdef \{ b \in B^\star \mid (a, b) \in X \}$.
  The first application of the dominated convergence theorem
  uses, like for the first marginal condition, the function
  $[b \in B^\star \mapsto \text{$e^\epsilon \cdot \mu_2(b)$ if $b \in B$, else $1$}]$
  as the dominating function (using Equation \eqref{eq:lord1}). The
  second application of the dominated convergence theorem uses
  $[a \in A^\star \mapsto \text{$\mu_1(a)$ if $a \in A$, else $0$}]$
  as the dominating function. Indeed, if $a \in A$, then
  \begin{align*}
    \sum_{b \in X\lside^a} \overline{\eta\lside^{\omega_n}}(a, b)
      &=    \sum_{b \in X\lside^a} \eta\lside^{\omega_n}(a, b)
       \leq \sum_{b \in B^\star} \eta\lside^{\omega_n}(a, b)
       =    \dfst(\eta\lside^{\omega_n})(a) = \mu_1^{\omega_n}(a) \leq \mu_1(a), \text{and}\\
    \sum_{b \in X\lside^a} \overline{\eta\lside}(\star, b)
      &= \sum_{b \in X\lside^a} 0 = 0 .
  \end{align*}

  \medskip

  Likewise, for $\overline{\eta\rside}[X]$, we have
  \begin{align*}
    \overline{\eta\rside}[X]
      &= \sum_{a, b \in X} \lim_{n \to \infty} \eta\rside^{\omega_n}(a, b)
       = \sum_{b \in B} \sum_{a \in X\rside^b} \lim_{n \to \infty} \eta\rside^{\omega_n}(a, b) \\
      &= \sum_{b \in B} \lim_{n \to \infty} \sum_{a \in X\rside^b} \eta\lside^{\omega_n}(a, b)
           \tag{dominated convergence} \\
      &= \lim_{n \to \infty} \sum_{b \in B} \sum_{a \in X\rside^b} \eta\lside^{\omega_n}(a, b)
           \tag{dominated convergence} \\
      &= \lim_{n \to \infty} \sum_{a, b \in X} \eta\rside^{\omega_n}(a, b)
       = \lim_{n \to \infty} \eta\rside^{\omega_n}(a, b)[X]
  \end{align*}
  where $X\rside^b \eqdef \{ a \in A^\star \mid (a, b) \in X \}$.
  Here, the first application of the dominated convergence theorem
  uses, as for the second marginal condition, the function
  $[a \in A^\star \mapsto
      \text{$\mu_1(a)$ if $a \in A$, else $1$}]$
  as the dominating function (using equation \eqref{eq:lord2}). The
  second application of the dominated convergence theorem uses
  $[b \in B^\star \mapsto
      \text{$\mu_2(b)$ if $b \in B$, else $0$}]$
  as the dominating function. Indeed, if $b \in B$, then
  \begin{align*}
    \sum_{a \in X\rside^b} \overline{\eta\rside^{\omega_n}}(a, b)
      &=    \sum_{a \in X\rside^b} \eta\rside^{\omega_n}(a, b)
       \leq \sum_{a \in A^\star} \eta\rside^{\omega_n}(a, b)
       =    \dsnd(\eta\rside^{\omega_n})(b) = \mu_2^{\omega_n}(b) \leq \mu_2(b), \text{and}\\
    \sum_{a \in X\rside^b} \overline{\eta\rside^{\omega_n}}(a, \star)
      &= \sum_{a \in X\rside^b} 0 = 0 .
  \end{align*}

  Hence, we can conclude the distance condition by taking limits:
  \begin{align*}
    \overline{\eta\lside}[X] - e^\epsilon \cdot \overline{\eta\rside}[X]
      &= \lim_{n \to \infty} \overline{\eta\lside^{\omega_n}}[X]
         - e^\epsilon \cdot \lim_{n \to \infty} \overline{\eta\rside^{\omega_n}}[X] \\
      &= \lim_{n \to \infty} \left(
           \overline{\eta\lside^{\omega_n}} - e^\epsilon \cdot
             \overline{\eta\rside^{\omega_n}}[X] \right)\\
      &\leq \lim_{n \to \infty} \delta_n = \delta . \qedhere
  \end{align*}
  
\end{proof}

This proof constructs witnesses to an approximate lifting relating two
distributions $(\mu_1, \mu_2)$ given a sequence of approximate liftings relating
pairs of finite restrictions of $\mu_1$ and $\mu_2$. Using essentially the same
argument, we can construct an approximate lifting relating $(\mu_1, \mu_2)$
given a sequence of approximate liftings relating pairs of distributions
converging to $\mu_1$ and $\mu_2$; the main difference is that a slightly more
general form of the dominated convergence theorem is needed (see~\citet[Lemma
5.1.7]{JHThesis} for details).

\section{%
\texorpdfstring
{Properties of $\star$-Liftings}
{Properties of *-Liftings}} \label{s:props}

Our main theorem can be used to show several natural properties of
$\star$-liftings. To begin, we can improve the mapping property from
Theorem~\ref{t:map:restr}, lifting the requirement that the maps must be
surjective.

\begin{lem} \label{l:alift:map}
  Let $\mu_1 \in \Dist(A_1)$, $\mu_2 \in \Dist(A_2)$,
  $f_1 : A_1 \to B_1$, $f_2 : A_2 \to B_2$ and $\rR$ a binary relation
  on $B_1$ and $B_2$. Let $\rS$ such that $a_1 \mathrel{\rS} a_2
    \iffdef f_1(a_1) \rR f_2(a_2)$. Then
  \[
    \dlift{f_1}(\mu_1) \aliftnew{\rR}{\epsilon,\delta} \dlift{f_2}(\mu_2)
    \iff
    \mu_1 \aliftnew{\rS}{\epsilon,\delta} \mu_2 .
  \]
\end{lem}

\proofatend
\begin{description}[style=unboxed,leftmargin=0cm]
\item[($\Longrightarrow$)]
  Assume that
  $\dlift{f_1}(\mu_1) \aliftnew{\rR}{\epsilon,\delta} \dlift{f_2}(\mu_2)$
  and let $X \subseteq A_1$. Then,
  \begin{align*}
    \mu_1[X]
      &\leq \mu_1[\inv{f_1}(f_1(X))] = \dlift{f_1}(\mu_1)[f_1(X)] \\
      &\leq e^{\epsilon} \cdot \dlift{f_2}(\mu_2)[\oR(f_1(X))] + \delta
         \tag{Theorem~\ref{t:alift:dom}} \\
      &= e^{\epsilon} \cdot \mu_2[\underbrace{%
           \inv{f_2}(\oR(f_1(X)))}_{\subseteq \rS(X)}]
           + \delta
       \leq e^{\epsilon} \cdot \mu_2[\rS(X)] + \delta .
  \end{align*}
  Hence, by Theorem~\ref{t:alift:dom},
  $\mu_1 \aliftnew{\rS}{\epsilon,\delta} \mu_2$.

\item[($\Longleftarrow$)]
  Assume that
  $\mu_1 \aliftnew{\rS}{\epsilon,\delta} \mu_2$
  and let $X \subseteq A_2$. Then,
  \begin{align*}
    \dlift{f_1}(\mu_1)[X]
      &= \mu_1[\inv{f_1}(X)] \\
      &\leq e^{\epsilon} \cdot \mu_2[\underbrace{%
           \rS(\inv{f_1}(X))}_{%
             \subseteq \inv{f_2}(\oR(X))}
         ] + \delta
        \tag{Theorem~\ref{t:alift:dom}}
       \leq e^{\epsilon} \cdot \dlift{f_2}(\mu_2)[\oR(X)] + \delta .
  \end{align*}
  Hence, by Theorem~\ref{t:alift:dom},
  $\dlift{f_1}(\mu_1) \aliftnew{\rR}{\epsilon,\delta} \dlift{f_2}(\mu_2)$.
  \qedhere
\end{description}
\endproofatend

Similarly, we can generalize the existing rules for up-to-bad reasoning (cf.\
\citet[Theorem 13]{BartheFGGHS16}), which restrict the post-condition to be
equality. There are two versions: the conditional event is either on the left
side, or the right side. Note that the resulting indices $\overline{\delta}$ are
different in the two cases. We write $\rcomp{\theta}$ for the complement of
$\theta$.

\begin{lem} \label{l:alift:utbl}
  Let $\mu_1 \in \Dist(A)$, $\mu_2 \in \Dist(B)$, $\theta \subseteq A$
  and $\oR \subseteq A \times B$. Assume that
  $\mu_1 \aliftnew{(\theta\lside \implies \rR)}{\epsilon,\delta} \mu_2$
  for some parameters $\epsilon, \delta \geq 0$.
  Then $\mu_1 \aliftnew{\rR}{\epsilon,\overline{\delta}} \mu_2$,
  where $\overline{\delta} \eqdef \delta + \mu_1[\rcomp{\theta}]$.
\end{lem}

\proofatend
  By Theorem~\ref{t:alift:dom}, it is sufficient to prove that
  \[ \mu_1[X] \leq e^{\epsilon} \cdot \mu_2[\oR(X)]
       + \mu_1[\rcomp{\theta}] + \delta \]
  for any $X \subseteq A$. By direct computation:
  \begin{align*}
    \mu_1[X] &= \mu_1[X \cap \theta]
      + \mu_1[X \cap \rcomp{\theta}]
       \leq \mu_1[X \cap \theta] + \mu_1[\rcomp{\theta}] \\
      &\leq e^{\epsilon} \cdot \mu_2[\underbrace{%
            (\theta\lside \implies \oR)(X \cap \theta)%
          }_{=\ \oR(X \cap \theta)\ \subseteq\ \oR(X)}]
        + \delta + \mu_1[\rcomp{\theta}] \\
      &\leq e^{\epsilon} \cdot \mu_2[\oR(X)]
              + \mu_1[\rcomp{\theta}] + \delta. \qedhere
  \end{align*}
\endproofatend

\begin{lem} \label{l:alift:utbr}
  Let $\mu_1 \in \Dist(A)$, $\mu_2 \in \Dist(B)$, $\theta \subseteq B$
  and $\oR \subseteq A \times B$. Assume that
  $\mu_1 \aliftnew{(\theta\rside \implies \rR)}{\epsilon,\delta} \mu_2$
  for some parameters $\epsilon, \delta \geq 0$.
  Then, $\mu_1 \aliftnew{\rR}{\epsilon,\overline{\delta}} \mu_2$,
  where $\overline{\delta} \eqdef \delta + e^\epsilon \cdot \mu_2[\rcomp{\theta}]$.
\end{lem}

\proofatend
  By Theorem~\ref{t:alift:dom}, it is sufficient to prove that
  \[ \mu_1[X] \leq e^{\epsilon} \cdot \mu_2[\oR(X)]
       + e^{\epsilon} \cdot \mu_2[\rcomp{\theta}] + \delta \]
  for any $X \subseteq A$. Let $X$ be such a set, then:
  \begin{align*}
    \mu_1[X]
      &\leq e^{\epsilon} \cdot \mu_2[(\theta\rside \implies \oR)(X)]
             + \delta \\
      &\leq e^{\epsilon} \cdot (
          \mu_2[(\theta\rside \implies \oR)(X) \cap \theta]
            + \mu_2[\rcomp{\theta}]) + \delta \\
      &\leq e^{\epsilon} \mu_2[\underbrace{%
                  (\theta\rside \implies \oR)(X) \cap \theta
                }_{\subseteq \oR(X) \cap \theta}]
              + e^{\epsilon} \cdot \mu_2[\rcomp{\theta}]
              + \delta \\
      &\leq e^{\epsilon} \mu_2[\oR(X)]
              + e^{\epsilon} \cdot \mu_2[\rcomp{\theta}]
              + \delta . \qedhere
  \end{align*}
\endproofatend

As a consequence, an approximately lifted relation can be conjuncted with a
one-sided predicate if the $\delta$ parameter is increased. This principle is
useful for constructing approximate liftings based on \emph{accuracy} bounds.
For instance, suppose that we have an approximate lifting of $\oR$ relating two
distributions, $\mu\lside$ and $\mu\rside$. If $\theta_{a}$ is an event that
happens with high probability in the first distribution $\mu\lside$---say, a
certain noise variable is at most $100$---we can incorporate $\theta_{a,
\lmark}$ into the approximate lifting by increasing the $\delta$ parameter by
the probability that $\theta_{a}$ \emph{fails} to hold in $\mu\lside$. When
reasoning in terms of approximate couplings, intuitively we can ``assume''
$\theta_{a}$ holds by ``paying'' with an increase in $\delta$. A similar
property holds for the second distribution $\mu\rside$.

We formalize these constructions with the following lemma.

\begin{lem} \label{l:onesided:and}
  Let $\mu_1 \in \Dist(A)$, $\mu_2 \in \Dist(B)$,
  $\theta_a \subseteq A$, $\theta_b \subseteq B$ and
  $\oR \subseteq A \times B$.
  Assume that
  $\mu_1 \aliftnew{\rR}{\epsilon,\delta} \mu_2$.
  Then, 
  $\mu_1 \aliftnew{(\theta_{a,\lmark} \cap \rR)}{\epsilon,\delta_a} \mu_2$
  and
  $\mu_1 \aliftnew{(\theta_{b,\rmark} \cap \rR)}{\epsilon,\delta_b} \mu_2$
  where $\delta_a \eqdef \delta + \mu_1[\rcomp{\theta_a}]$ and
  $\delta_b \eqdef \delta + e^{\epsilon} \cdot \mu_2[\rcomp{\theta_b}]$.
\end{lem}

\proofatend
  From
  $\mu_1 \aliftnew{\rR}{\epsilon,\delta} \mu_2$
  and Lemma~\ref{l:alift:basics}, we have
  $\mu_1 \aliftnew{\rS}{\epsilon,\delta} \mu_2$
  where
  $\oS \eqdef \theta_{a,\lmark} \implies \theta_{a,\lmark} \cap \oR$.
  Hence, by Lemma~\ref{l:alift:utbl}, we obtain
  $\mu_1 \aliftnew{(\theta_{a,\lmark} \cap \oR)}{\epsilon,\delta_a} \mu_2$.
  Using similar reasoning with
  $\theta_{b,\rmark} \implies \theta_{b,\rmark} \cap \oR$
  and Lemma~\ref{l:alift:utbr}, we have
  $\mu_1 \aliftnew{(\theta_{b,\rmark} \cap \oR)}{\epsilon,\delta_b} \mu_2$.
\endproofatend

$\star$-liftings also support a significant generalization of optimal subset
coupling. Unlike the known construction for $2$-liftings
(Theorem~\ref{t:subset:opt}), the two subsets need not be nested, and either
subset may be the entire domain. Furthermore, the distributions $\mu_1, \mu_2$
need not be the same, or even have the same domain. Finally, the equivalence is
valid for any parameters $(\epsilon, \delta)$, not just $\delta = 0$.

\begin{thm} \label{t:subset:star}
  Let $\mu_1 \in \Dist(A_1)$, $\mu_2 \in \Dist(A_2)$ and consider two subsets $P_1
  \subseteq A_1, P_2 \subseteq A_2$. Then, we have the following equivalence:
  \[
    \sP{\mu_1}{P_1} \leq e^\epsilon \cdot \sP{\mu_2}{P_2} + \delta
    \quad\text{and}\quad
    \sP{\mu_1}{A_1 \setminus P_1} \leq e^\epsilon \cdot \sP{\mu_2}{A_2 \setminus P_2} + \delta
    \iff
    \mu_1 \aliftnew{\rR}{\epsilon, \delta} \mu_2 ,
  \]
  where
  $a_1 \mathrel{\rR} a_2
    \iffdef a_1 \in P_1 \iff a_2 \in P_2$.
\end{thm}

\begin{proof}
  Immediate by Theorem~\ref{t:alift:dom}.
\end{proof}

We can then recover the existing notion of optimal subset coupling
\citep{BartheFGGHS16} for $\star$-liftings as a special case.

\begin{cor}[\citet{BartheFGGHS16}] \label{c:subset:star}
  Let $\mu \in \Dist(A)$ and consider two nested subsets $P_2 \subseteq P_1
  \subseteq A$.  Then, we have the following equivalence:
  \[
    \sP{\mu}{P_1} \leq e^\epsilon \cdot \sP{\mu}{P_2}
    \iff
    \mu_1 \aliftnew{\rR}{\epsilon, 0} \mu_2 ,
  \]
  where
  $a_1 \mathrel{\rR} a_2
    \iffdef a_1 \in P_1 \iff a_2 \in P_2$.
\end{cor}
\begin{proof}
  Immediate by Theorem~\ref{t:subset:star}, noting that
  \[
    \sP{\mu}{A \setminus P_1} \leq e^\epsilon \cdot \sP{\mu}{A \setminus P_2}
  \]
  is automatic since $P_2 \subseteq P_1$ implies $\sP{\mu}{A \setminus P_1} \leq
  \sP{\mu}{A \setminus P_2}$. Note that there is no longer a need for $P_1$ to
  be a strict subset of $A$.
\end{proof}

Finally, we can directly extend known composition theorems from differential
privacy to $\star$-liftings. This connection is quite useful for transferring
existing composition results from the privacy literature to approximate
liftings. We first define a general template describing how the privacy
parameters $\epsilon, \delta$ decay under sequential composition.

\begin{defi} \label{d:comp:rule}
  Let $\RR^{\geq 0}_2 \eqdef \RR^{\geq 0} \times \RR^{\geq 0}$ and let
  $(\RR^{\geq 0}_2)^*$ be the set of finite sequences over pairs of non-negative
  reals. A map $r : (\RR^{\geq 0}_2)^* \to \RR^{\geq 0}_2$ is a
  \emph{DP-composition rule} if for all sets $A, D$, adjacency relations $\phi
  \subseteq D \times D$, and families of functions $\{ f_i : D \times A \to
  \Dist(A) \}_{i < n}$, the following implication holds: if for every initial
  value $a \in A$ and $i < n$, $f_i(-,a) : D \to \Dist(A)$ is $(\epsilon_i,
  \delta_i)$-differentially private w.r.t.\ $\phi$, then $F(-,a)$ is
  $(\epsilon^*, \delta^*)$-differentially private w.r.t.\ $\phi$ and any initial
  value $a \in A$ where $F : (d, a) \mapsto ( \bigcirc_{i < n}
  \lift{(f_i(d,-))})(\dunit{a})$ is the $n$-fold composition of the functions
  $[f_i]_{i<n}$ and $(\epsilon^*, \delta^*) \eqdef r([(\epsilon_i, \delta_i)]_{i < n})$.
\end{defi}

\begin{lem} \label{l:star:comp}
  Let $r : (\RR^{\geq 0}_2)^* \to \RR^{\geq 0}_2$ be a DP-composition rule.
  Let $n \in \NN$ and assume given two families of sets
  $\{A_i\}_{i \leq n}$ and $\{B_i\}_{i \leq n}$, together with a
  family of binary relations
  $\{ \oR(i) \subseteq A_i \times B_i \}_{i \leq n}$.
  Let
  $\{ g_i : A_i \to \Dist(A_{i+1}) \}_{i < n}$ and
  $\{ h_i : B_i \to \Dist(B_{i+1}) \}_{i < n}$ be two families of functions s.t.
  for all $i < n$ and $(a, b) \in \oR(i)$, we have:
  \begin{enumerate}
  \item
    $g_i(a) \mathrel{\aliftnew{\oR(i + 1)}{\epsilon_i, \delta_i}}
    h_i(b)$ for some parameters $\epsilon_i, \delta_i \geq 0$, and
  \item
    $g_i(a)$ and $h_i(b)$ are proper distributions.
  \end{enumerate}
  Then for $(a_0, b_0) \in \oR(0)$, there exists a $\star$-lifting
  \[
    G(a_0) \mathrel{\aliftnew{\oR(n)}{\epsilon^*, \delta^*}} H(b_0) 
  \]
  where $G : A_0 \to \Dist(A_n)$ and $H : B_0 \to \Dist(B_n)$ are the $n$-fold
  compositions of $[ g_i ]_{i \leq n}$ and $[ h_i ]_{i \leq n}$
  respectively---\ie $G(a) \eqdef (\bigcirc_{i < n} \lift{g_i})(\dunit{a})$ and
  $H(b) \eqdef (\bigcirc_{i < n} \lift{h_i})(\dunit{b})$---and $(\epsilon^*,
  \delta^*) \eqdef r([(\epsilon_i, \delta_i)]_{i < n})$.
\end{lem}
\begin{proof}
  We assume that $A_i = A$ is the same for all $i$, and $B_i = B$ is the same
  for all $i$. This is without loss of generality, since when $A_i$ and $B_i$
  vary with $i$ we may work with the disjoint unions $\sqcup_i A_i$ and
  $\sqcup_i B_i$ by restricting each $\oR(i)$ to only relate pairs that are both
  in the $i$-th component. Define $D = \BB = \{ \mathit{tt}, \mathit{ff} \}$ and
  $\phi = \{ (\mathit{tt}, \mathit{ff}) \} \subseteq D \times D$.
  
  For every $i < n$ and $(a_i, b_i) \in \oR(i)$, the definition of
  $\star$-lifting gives two distributions $\mu\lside[a_i, b_i], \mu\rside[a_i,
  b_i]$ witnessing $g_i(a) \mathrel{\aliftnew{\oR(i + 1)}{\epsilon_i,
  \delta_i}}$.  We regard both witness distributions as elements of
  $\Dist(A^\star \times B^\star)$ via the evident embeddings. We define the maps
  $f_i : D \times (A^\star \times B^\star) \to \Dist(A^\star \times B^\star)$ by
  cases:
  \begin{align*}
    f_i(\mathit{tt}, (a_i, b_i)) &= \mu\lside[a_i, b_i] \\
    f_i(\mathit{ff}, (a_i, b_i)) &= \mu\rside[a_i, b_i] \\
    f_i(-, (a_i, \star)) &= g_i(a_i) \times \dunit{\star} \\
    f_i(-, (\star, b_i)) &= \dunit{\star} \times h_i(b_i) \\
    f_i(-, (\star, \star)) &= \dunit{(\star, \star)}
  \end{align*}
  where $\times$ denotes the product distribution and $(a_i, b_i) \in \oR(i)$;
  otherwise, $f_i(-, (a, b)) = 0$.
  
  Now for all $(a, b) \in A^\star \times B^\star$, the map $f_i(-, (a, b)) : D
  \to \Dist(A^\star \times B^\star)$ is $(\epsilon_i, \delta_i)$-differentially
  private with respect to $\phi$ by the distance property on $\mu\lside[a_i,
  b_i]$ and $\mu\rside[a_i, b_i]$ (and by definition when $(a_i, b_i) \notin
  \oR(i)$). Hence, the DP-composition rule implies that $F : (d, (a, b)) \mapsto
  ( \bigcirc_{i < n} \lift{(f_i(d,-))})(\dunit{(a, b)})$ is $(\epsilon^*,
  \delta^*)$-differentially private with respect to $\phi$ for any $(a, b) \in
  A^\star \times B^\star$.  For any $(a_0, b_0) \in \oR(0)$, we claim that
  that $F(\mathit{tt}, (a_0, b_0))$ and $F(\mathit{ff}, (a_0, b_0))$ witness the
  desired approximate lifting
  \[
    G(a_0) \mathrel{\aliftnew{\oR(n)}{\epsilon^*, \delta^*}} H(b_0) .
  \]
  The support and marginal conditions are not hard to show, and the distance
  condition follows from differential privacy of $F(-, (a_0, b_0)) : D \to
  \Dist(A^\star \times B^\star)$.
\end{proof}

Some of the more sophisticated composition results from differential
privacy---for instance, the advanced composition theorem by \citet{DRV10}---do
not apply to arbitrary adjacency relations $\phi$, but only \emph{symmetric}
relations. Lemma~\ref{l:star:comp} cannot lift such theorems to composition
principles for approximate liftings. In Section~\ref{s:sym} we will remedy this
problem by working with a symmetric version of $\star$-lifting.

\section{Comparison with Prior Approximate Liftings} \label{s:comp}
Now that we have seen $\star$-liftings, we briefly consider other definitions of
approximate liftings. We have already seen $2$-liftings, which involve two
witnesses (Definition~\ref{d:2lift}). Evidently, $\star$-liftings strictly
generalize $2$-liftings.

\begin{thm} \label{l:incl:star}
  For all binary relations $\rR$ over $A$ and $B$ and parameters
  $\epsilon, \delta \geq 0$, we have
  $
    \alifticalp{\rR}{\epsilon, \delta} \subseteq \aliftnew{\rR}{\epsilon, \delta} .
  $
  There exist relations and parameters where the inclusion is strict.
\end{thm}

\begin{proof}
  The inclusion $\alifticalp{\rR}{\epsilon, \delta} \subseteq
  \aliftnew{\rR}{\epsilon, \delta}$ is immediate.  We have a strict inclusion
  $\alifticalp{\rR}{\epsilon, \delta} \subsetneq \aliftnew{\rR}{ \epsilon,
    \delta}$ even for $\delta = 0$ by considering the optimal subset coupling
  from Theorem~\ref{t:subset:opt}. Consider a distribution $\mu$ over set $A$,
  and let $P_1 \subseteq P_2 = A$.  There is an $(\epsilon, 0)$-approximate
  $\star$-lifting (by Theorem~\ref{t:subset:star}), but a $(\epsilon,
  0)$-approximate $2$-lifting does not exist if $\mu$ has non-zero mass outside
  of $P_1$: the first witness $\mu\lside$ must place non-zero mass at
  $(a_1,a_2)$ with $a_1 \notin P_1$ in order to have $\dfst(\mu\lside) = \mu$,
  but we must have $a_2 \notin P_2$ for the support requirement, and there is no
  such $a_2$.
\end{proof}

We can also compare $\star$-liftings with the original definitions of
$(\epsilon,\delta)$-approximate lifting by \citet{BartheKOZ13}. They introduce
two notions, a symmetric lifting and an asymmetric lifting, each using a single
witness distribution. We will focus on the asymmetric version here, and return
to the symmetric version in Section~\ref{s:sym}.

\begin{defi}[\citet{BartheKOZ13}]
  Let $\mu_1 \in \Dist(A)$ and $\mu_2 \in \Dist(B)$ be sub-distributions,
  $\epsilon, \delta \in \RR^{\geq 0}$ and $\rR$ be a binary relation over $A
$ and $B$. An $(\epsilon, \delta)$-\emph{approximate} $1$-\emph{lifting}
  of $\mu_1$ and $\mu_2$ for $\rR$ is a sub-distribution $\mu \in \Dist(A
  \times B)$ s.t.
  \begin{enumerate}
    \item $\dfst(\mu) \leq \mu_1$ and $\dsnd(\mu) \leq \mu_2$;
    \item $\Delta_\epsilon(\mu_1, \dfst(\mu)) \leq \delta$; and
    \item $\supp(\mu) \subseteq \rR$.
  \end{enumerate}
  In the first point we take the point-wise order on sub-distributions: if $\mu$
  and $\mu'$ are sub-distributions over $X$, then $\mu \leq \mu'$ when $\mu(x)
  \leq \mu'(x)$ for all $x \in X$. We will write
  $
    \mu_1 \alifttoplas{\rR}{\epsilon, \delta} \mu_2
  $
  if there exists an $(\epsilon, \delta)$-approximate $1$-lifting of $\mu_1
$ and $\mu_2$ for $\rR$; the superscript $\cdot^{(1)}$ indicates that
  there is one witness for this lifting.
\end{defi}

$1$-liftings bear a close resemblance to \emph{probabilistic couplings} from
probability theory, which also have a single witness.  However, $1$-liftings are
more awkward to manipulate and less well-understood theoretically than their
$2$-lifting cousins---basic properties such as mapping (Lemma~\ref{l:alift:map})
are not known to hold; the subset coupling (Theorem~\ref{t:subset:opt}) is not
known to exist.  Somewhat surprisingly, $1$-liftings are equivalent to
$\star$-liftings and hence by Theorem~\ref{t:alift:dom}, also to Sato's
approximate lifting.

\begin{thm} \label{t:1-star}
  For all binary relations $\rR$ over $A$ and $B$ and parameters
  $\epsilon, \delta \geq 0$, we have
  $
    \alifttoplas{\rR}{\epsilon, \delta} = \aliftnew{\rR}{\epsilon, \delta} .
  $
\end{thm}
\proofatend
  Suppose that $(\mu_L, \mu_R)$ are witnesses to $\mu_1 \aliftnew{\rR}{\epsilon,
    \delta} \mu_2$. Define the witness $\eta \in \Dist(A \times B)$ as the
  point-wise minimum: $\eta(a, b) = \min(\mu_L(a, b), \mu_R(a, b))$. We
  will show that $\eta$ is a witness to $\mu_1
  \alifttoplas{\rR}{\epsilon,\delta} \mu_2$.

  The support condition follows from the support condition for $(\mu_L, \mu_R)$.
  The marginal conditions $\dfst(\eta) \leq \mu_1$ and $\dsnd(\eta) \leq \mu_2$
  also follow by the marginal conditions for $(\mu_L, \mu_R)$. The only thing to
  check is the distance condition. By the distance condition on $(\mu_L,
  \mu_R)$, there exist non-negative values $\delta(a, b)$ such that
  \[
    \mu_L(a, b) \leq \exp(\epsilon) \mu_R(a, b) + \delta(a, b)
  \]
  and $\sum_{a, b} \delta(a,b) \leq \delta$. So, $\mu_R(a, b) \geq
  \exp(-\epsilon) (\mu_L(a, b) - \delta(a, b))$.  Now let $S \subseteq A$ be
  any subset. Then:
  \begin{align*}
    \mu_1(S) - \exp(\epsilon) \dfst(\eta)(S)
    &= \sum_{a \in S} \mu_1(a) - \exp(\epsilon) \sum_{b \in B} \min(\mu_L(a, b), \mu_R(a, b)) \\
    &\leq \sum_{a \in S} \mu_1(a) - \exp(\epsilon)
    \sum_{b \in B} \exp(-\epsilon) (\mu_L(a, b) - \delta(a, b)) \\
    &= \sum_{a \in S, b \in B} \delta(a, b) \leq \delta .
  \end{align*}
  Thus, $\eta$ witnesses $\mu_1 \alifttoplas{\rR}{\epsilon, \delta} \mu_2$, so
  $\aliftnew{\rR}{\epsilon, \delta} \subseteq \alifttoplas{\rR}{\epsilon,
    \delta}$.

  The other direction is more interesting. Let $\eta \in \Dist(A \times B)$ be
  the witness for $\alifttoplas{\rR}{\epsilon, \delta}$.  By the distance
  condition $\Delta_{\epsilon}(\mu_1,\dfst \eta) \leq \delta$, there exist
  non-negative values $\delta(a)$ such that
  \[
    \mu_1(a) \leq \exp(\epsilon) \dfst \eta(a) + \delta(a)
  \]
  with equality when $\delta(a)$ is strictly positive, and $\sum_{a \in A}
  \delta(a) \leq \delta$. Define two witnesses $\mu_L \in \Dist(A \times
  B^\star), \mu_R \in \Dist(A^\star \times B)$ as follows:
  \begin{align*}
    \mu_L(a, b) &=
    \begin{cases}
      \eta(a, b) \cdot \frac{\mu_1(a) - \delta(a)}{\dfst\eta(a)}
      &: b \neq \star \\
      \mu_1(a) - \sum_{b \in B} \mu_L(a, b)
      &: b = \star
    \end{cases}
    \\
    \mu_R(a, b) &=
    \begin{cases}
      \eta(a, b) &: a \neq \star \\
      \mu_2(b) - \sum_{a \in A} \mu_R(a, b)
      &: a = \star .
    \end{cases}
  \end{align*}
  (As usual, if any denominator is zero, we take the probability to be zero as
  well.)

  The support condition follows from the support condition of $\eta$. The
  marginal conditions hold by definition. Note that all probabilities are
  non-negative. For $\mu_L$, note that if $\delta(a) > 0$ then $\mu_1(a) -
  \delta(a) = \exp(\epsilon) \dfst \eta(a) \geq 0$ and hence
  \[
    \mu_L(a, \star) = \mu_1(a) - \delta(a) \geq 0 .
  \]
  assuming $\dfst\eta(a) > 0$; if $\dfst\eta(a) = 0$ then $\mu_L(a, \star) = 0$.
  For $\mu_R$, non-negativity holds because $\dsnd \eta \leq \mu_2$.
  
  We show the distance bound.  Note that when $a, b \neq \star$, by definition
  $\mu_L(a, b)$ and $\mu_R(a, b)$ are both strictly positive or both equal to
  zero, and $\eta(a, b)$ is strictly positive or equal to zero accordingly.  If
  $\mu_L(a, b), \mu_R(a, b), \eta(a, b)$ are all strictly positive, then we know
  \[
    \frac{\mu_L(a, b)}{\eta(a, b)}
    = \frac{\mu_1(a) - \delta(a)}{\dfst\eta(a)} \leq \exp(\epsilon) .
  \]
  Thus we always have
  \[
    \mu_L(a, b) \leq \exp(\epsilon) \eta(a, b) = \exp(\epsilon) \mu_R(a, b) .
  \]
  We can also bound the mass on points $(a, \star)$. Let $S \subseteq A$ be
  any subset. Then:
  \begin{align*}
    \overline{\mu_L}(S \times \{ \star \}) &= \sum_{a \in S} \mu_1(a)
    - \mu_1(a) \sum_{b \in B} \frac{\eta(a, b)}{\dfst \eta(a)}
    + \delta(a) \sum_{b \in B} \frac{\eta(a, b)}{\dfst \eta(a)} \\
    &= \mu_1(S) - \mu_1(S) + \delta(S)
    \leq \exp(\epsilon) \overline{\mu_R}(S \times \{ \star \}) + \delta .
  \end{align*}
  So $\Delta_{\epsilon}(\overline{\mu_L},\overline{\mu_R}) \leq \delta$ as
  desired, and we have witnesses to $\mu_1 \aliftnew{\rR}{\epsilon, \delta}
  \mu_2$. Hence, $\alifttoplas{\rR}{\epsilon, \delta} \subseteq
  \aliftnew{\rR}{\epsilon, \delta}$.
\endproofatend

\section{%
  \texorpdfstring
  {Symmetric $\star$-Lifting}
  {Symmetric *-Lifting}} \label{s:sym}

The approximate liftings we have considered so far are all \emph{asymmetric}.
For instance, the approximate lifting $\mu_1 \aliftnew{\rR}{\epsilon, \delta}
\mu_2$ may not imply the lifting $\mu_2 \aliftnew{(\rR^{-1})}{\epsilon, \delta}
\mu_1$. Given witnesses $(\mu_L, \mu_R)$ to the first lifting, we may consider
the witnesses $(\nu_L, \nu_R) = (\mu_R^\top, \mu_L^\top)$ where the
transpose map $(-)^\top : \Dist(A \times B) \to \Dist(B \times A)$ is defined in
the obvious way. Then $(\nu_L, \nu_R)$ almost witness the second lifting---the
marginal and support conditions holds, but the distance bound is in the wrong
direction:
\[
  \Delta_\epsilon(\nu_R, \nu_L)
  = \Delta_\epsilon(\mu_L^\top, \mu_R^\top) 
  = \Delta_\epsilon(\mu_L, \mu_R)
  \leq \delta .
\]
In general, we cannot bound $\Delta_\epsilon(\nu_L, \nu_R)$ and the symmetric
lifting $\mu_2 \aliftnew{(\rR^{-1})}{\epsilon, \delta} \mu_1$ may not hold.
To recover symmetry, we can define a symmetric version of $\star$-lifting.
\begin{defi}[Symmetric $\star$-lifting]
  Let $\mu_1 \in \Dist(A)$ and $\mu_2 \in \Dist(B)$ be sub-distributions,
  $\epsilon, \delta \in \RR^{\geq 0}$ and $\rR$ be a binary relation over $A
$ and $B$. An $(\epsilon, \delta)$-\emph{approximate symmetric}
  $\star$-\emph{lifting} of $\mu_1$ and $\mu_2$ for $\rR$ is a pair of
  sub-distributions
  $\eta\lside \in \Dist(A \times B^\star)$ and
  $\eta\rside \in \Dist(A^\star \times B)$ s.t.
  \begin{enumerate}
  \item $\dfst(\eta\lside) = \mu_1$ and $\dsnd(\eta\rside) = \mu_2$;
  \item
    $\supp({\eta\lside}_{|A \times B}) ,
     \supp({\eta\rside}_{|A \times B}) \subseteq \oR$; and
  \item $\Delta_{\epsilon}(
           \overline{\eta\lside},
           \overline{\eta\rside}) \leq \delta,
           \Delta_{\epsilon}(
           \overline{\eta\rside},
           \overline{\eta\lside}) \leq \delta$,
  where $\overline{\eta_{\bullet}}$ is the canonical lifting of
  $\eta_{\bullet}$ to $A^\star \times B^\star$. 
  \end{enumerate}
  We write
  $\mu_1 \symliftnew{R}{\epsilon, \delta} \mu_2$
  if there exists an $(\epsilon, \delta)$-approximate symmetric lifting of
  $\mu_1$ and $\mu_2$ for $\rR$.
\end{defi}

Symmetric $\star$-lifting is a special case of $\star$-lifting that can capture
differential privacy under when the adjacency relation $\phi$ is
\emph{symmetric}: a probabilistic computation $M : A \to \Dist(B)$ is
$(\epsilon, \delta)$-differentially private if and only if for every two
adjacent inputs $a \mathrel{\phi} a'$, there is an approximate lifting of the
equality relation: $M(a) \symliftnew{(=)}{\epsilon, \delta} M(a')$.
Unfortunately, the more advanced properties in Section~\ref{s:props} do not all
hold when moving to symmetric liftings. However, we can show that symmetric
$\star$-liftings are equivalent to the symmetric version of $1$-witness lifting
proposed by \citet{BartheKOZ13}.

\begin{defi}[\citet{BartheKOZ13}]
  Let $\mu_1 \in \Dist(A)$ and $\mu_2 \in \Dist(B)$ be sub-distributions,
  $\epsilon, \delta \in \RR^{\geq 0}$ and $\rR$ be a binary relation over $A
$ and $B$. An $(\epsilon, \delta)$-\emph{approximate symmetric}
  $1$-\emph{lifting} of $\mu_1$ and $\mu_2$ for $\rR$ is a
  sub-distribution $\mu \in \Dist(A \times B)$ s.t.
  \begin{enumerate}
    \item $\dfst(\mu) \leq \mu_1$ and $\dsnd(\mu) \leq \mu_2$;
    \item $\Delta_\epsilon(\mu_1, \dfst(\mu)) \leq \delta$ and
      $\Delta_\epsilon(\mu_2, \dsnd(\mu)) \leq \delta$; and
    \item $\supp(\mu) \subseteq \rR$.
  \end{enumerate}
  We will write
  $
    \mu_1 \symlifttoplas{\rR}{\epsilon, \delta} \mu_2
  $
  if there exists an $(\epsilon, \delta)$-approximate symmetric $1$-lifting of
  $\mu_1$ and $\mu_2$ for $\rR$; the superscript $\cdot^{(1)}$ indicates
  that there is one witness for this lifting.
\end{defi}

\begin{thm}[cf.\ the asymmetric result Theorem~\ref{t:1-star}] \label{t:sym-1-star}
  For all binary relations $\rR$ over $A$ and $B$ and parameters
  $\epsilon, \delta \geq 0$, we have
  $
    \symlifttoplas{\rR}{\epsilon, \delta} = \symliftnew{\rR}{\epsilon, \delta} .
  $
\end{thm}
\proofatend
  Suppose that $(\mu_L, \mu_R)$ are witnesses to $\mu_1
  \symliftnew{\rR}{\epsilon, \delta} \mu_2$. Define the witness $\eta \in
  \Dist(A \times B)$ as the point-wise minimum: $\eta(a, b) =
  \min(\mu_L(a, b), \mu_R(a, b))$. We will show that $\eta$ is a witness to
  $\mu_1 \symlifttoplas{\rR}{\epsilon, \delta} \mu_2$.

  The support condition follows from the support condition for $(\mu_L, \mu_R)$.
  The marginal conditions $\dfst(\eta) \leq \mu_1$ and $\dsnd(\eta) \leq \mu_2$
  also follow by the marginal conditions for $(\mu_L, \mu_R)$. The only thing to
  check is the distance condition. By the distance condition on $(\mu_L,
  \mu_R)$, there exist non-negative values $\delta(a, b)$ such that
  \[
    \mu_L(a, b) \leq \exp(\epsilon) \mu_R(a, b) + \delta(a, b)
  \]
  and $\sum_{a, b} \delta(a,b) \leq \delta$. So, $\mu_R(a, b) \geq
  \exp(-\epsilon) (\mu_L(a, b) - \delta(a, b))$. Similarly, there are
  non-negative values $\delta'(a, b)$ such that
  \[
    \mu_R(a, b) \leq \exp(\epsilon) \mu_L(a, b) + \delta'(a, b)
  \]
  and $\sum_{a, b} \delta'(a,b) \leq \delta$. So, $\mu_L(a, b) \geq
  \exp(-\epsilon) (\mu_R(a, b) - \delta'(a, b))$.
  
  Now let $S \subseteq A$ be any subset. Then:
  \begin{align*}
    \mu_1(S) - \exp(\epsilon) \dfst(\eta)(S)
    &= \sum_{a \in S} \mu_1(a) - \exp(\epsilon) \sum_{b \in B} \min(\mu_L(a, b), \mu_R(a, b)) \\
    &\leq \sum_{a \in S} \mu_1(a) - \exp(\epsilon)
    \sum_{b \in B} \exp(-\epsilon) (\mu_L(a, b) - \delta(a, b)) \\
    &= \sum_{a \in S, b \in B} \delta(a, b) \leq \delta .
  \end{align*}
  The other marginal is similar. For any subset $T \subseteq B$ we have
  \begin{align*}
    \mu_2(T) - \exp(\epsilon) \dsnd(\eta)(T)
    &= \sum_{b \in T} \mu_2(b) - \exp(\epsilon) \sum_{a \in A} \min(\mu_L(a, b), \mu_R(a, b)) \\
    &\leq \sum_{b \in T} \mu_2(b) - \exp(\epsilon)
    \sum_{a \in A} \exp(-\epsilon) (\mu_R(a, b) - \delta'(a, b)) \\
    &= \sum_{b \in T, a \in A} \delta'(a, b) \leq \delta .
  \end{align*}
  Thus, $\eta$ witnesses $\mu_1 \symlifttoplas{\rR}{\epsilon, \delta} \mu_2$.

  The other direction is more interesting. Let $\eta \in \Dist(A \times B)$ be
  the single witness to $\symlifttoplas{\rR}{\epsilon,\delta}$.  By the
  distance conditions $\Delta_{\epsilon}(\mu_1,\dfst \eta) \leq \delta$ and
  $\Delta_{\epsilon}(\mu_2,\dsnd \eta) \leq \delta$, there exist non-negative
  values $\delta(a)$ and $\delta'(b)$ such that
  \begin{align*}
    \mu_1(a) &\leq \exp(\epsilon) \dfst \eta(a) + \delta(a) \\
    \mu_2(b) &\leq \exp(\epsilon) \dsnd \eta(b) + \delta'(b) ,
  \end{align*}
  there is equality when $\delta(a)$ or $\delta'(b)$ are strictly positive, and
  both $\sum_{a \in A} \delta(a)$ and $\sum_{b \in B} \delta'(b)$ are at
  most $\delta$. Define two witnesses $\mu_L \in \Dist(A \times B^\star), \mu_R
  \in \Dist(A^\star \times B)$ as follows:
  \begin{align*}
    \mu_L(a, b) &=
    \begin{cases}
      \eta(a, b) \cdot \frac{\mu_1(a) - \delta(a)}{\dfst\eta(a)}
      &: b \neq \star \\
      \mu_1(a) - \sum_{b \in B} \mu_L(a, b)
      &: b = \star
    \end{cases}
    \\
    \mu_R(a, b) &=
    \begin{cases}
      \eta(a, b) \cdot \frac{\mu_2(b) - \delta'(b)}{\dsnd\eta(b)}
      &: a \neq \star \\
      \mu_2(b) - \sum_{a \in A} \mu_R(a, b)
      &: a = \star .
    \end{cases}
  \end{align*}
  (As usual, if any denominator is zero, we take the probability to be zero as
  well.)
  
  The support condition follows from the support condition of $\eta$. The
  marginal conditions hold by definition. Note that all probabilities are
  non-negative. For instance in $\mu_L$, note that if $\delta(a) > 0$ then
  $\mu_1(a) - \delta(a) = \exp(\epsilon) \dfst \eta(a) \geq 0$ and hence
  \[
    \mu_L(a, \star) = \mu_1(a) - \delta(a) \geq 0 .
  \]
  assuming $\dfst\eta(a) > 0$; if $\dfst\eta(a) = 0$ then $\mu_L(a, \star) = 0$.
  A similar argument shows that $\mu_R$ is non-negative.
  
  So, it remains to check the distance bounds.  Note that when $a, b \neq
  \star$, by definition $\mu_L(a, b)$ and $\mu_R(a, b)$ are both strictly
  positive or both equal to zero, and $\eta(a, b)$ is strictly positive or equal
  to zero accordingly. If $\mu_L(a, b), \mu_R(a, b), \eta(a, b)$ are all
  strictly positive, then we know
  \begin{align*}
    \frac{\mu_L(a, b)}{\eta(a, b)}
    &= \frac{\mu_1(a) - \delta(a)}{\dfst\eta(a)}
    \leq \exp(\epsilon) \\
    \frac{\mu_R(a, b)}{\eta(a, b)}
    &= \frac{\mu_2(b) - \delta'(b)}{\dsnd\eta(b)}
    \leq \exp(\epsilon) .
  \end{align*}
  We can also lower bound the ratios:
  \begin{align*}
    \frac{\mu_L(a, b)}{\eta(a, b)}
    &= \frac{\mu_1(a) - \delta(a)}{\dfst\eta(a)}
    \geq 1 \\
    \frac{\mu_R(a, b)}{\eta(a, b)}
    &= \frac{\mu_2(b) - \delta'(b)}{\dsnd\eta(b)}
    \geq 1 ;
  \end{align*}
  for instance when $\delta(a) > 0$ then the ratio is exactly equal to
  $\exp(\epsilon) \geq 1$, and when $\delta(a) = 0$ then the ratio is at least
  $1$ by the marginal property $\dfst\eta \leq \mu_1$.
  So we have $\mu_L(a, b)/\eta(a, b)$ and $\mu_R(a, b)/\eta(a, b)$ in $[1,
  \exp(\epsilon)]$ when all distributions are strictly positive. Thus we always
  have
  \begin{align*}
    \mu_L(a, b) &\leq \exp(\epsilon) \mu_R(a, b) \\
    \mu_R(a, b) &\leq \exp(\epsilon) \mu_L(a, b) .
  \end{align*}
  We can also bound the mass on points $(a, \star)$. Let $S \subseteq A$ be
  any subset. $\overline{\mu_R}(S \times \{ \star \}) \leq \exp(\epsilon)
  \overline{\mu_L}(S \times \{ \star \}) + \delta$ is clear. For the other
  direction:
  \begin{align*}
    \overline{\mu_L}(S \times \{ \star \}) &= \sum_{a \in S} \mu_1(a)
    - \mu_1(a) \sum_{b \in B} \frac{\eta(a, b)}{\dfst \eta(a)}
    + \delta(a) \sum_{b \in B} \frac{\eta(a, b)}{\dfst \eta(a)} \\
    &= \mu_1(S) - \mu_1(S) + \delta(S)
    \leq \exp(\epsilon) \overline{\mu_R}(S \times \{ \star \}) + \delta .
  \end{align*}
  The mass at points $(\star, b)$ can be bounded in a similar way. Let $T
  \subseteq B$ be any subset. Then, $\overline{\mu_L}(\{ \star \} \times T) \leq
  \exp(\epsilon) \overline{\mu_R}(\{ \star \} \times T) + \delta$ is clear. For
  the other direction:
  \begin{align*}
    \overline{\mu_R}(\{ \star \} \times T) &= \sum_{b \in T} \mu_2(b)
    - \mu_2(b) \sum_{a \in A} \frac{\eta(a, b)}{\dsnd \eta(b)}
    + \delta'(b) \sum_{a \in A} \frac{\eta(a, b)}{\dsnd \eta(b)} \\
    &= \mu_2(T) - \mu_2(T) + \delta'(T)
    \leq \exp(\epsilon) \overline{\mu_L}(\{ \star \} \times T) + \delta .
  \end{align*}
  So $\Delta_{\epsilon}(\overline{\mu_L},\overline{\mu_R}) \leq \delta$ and
  $\Delta_{\epsilon}(\overline{\mu_R},\overline{\mu_L}) \leq \delta$ so we have
  witnesses to $\mu_1 \symliftnew{\rR}{\epsilon, \delta} \mu_2$. Hence,
  $\symlifttoplas{\rR}{\epsilon, \delta} = \symliftnew{\rR}{\epsilon, \delta}$.
\endproofatend

The main use of symmetric approximate liftings is to support richer composition
results that only apply to symmetric adjacency relations.

\begin{defi} \label{d:sym:comp:rule}
  Let $\RR^{\geq 0}_2 \eqdef \RR^{\geq 0} \times \RR^{\geq 0}$ and let
  $(\RR^{\geq 0}_2)^*$ be the set of finite sequences over pairs of non-negative
  reals. A map $r : (\RR^{\geq 0}_2)^* \to \RR^{\geq 0}_2$ is a
  \emph{symmetric DP-composition rule} if for all sets $A, D$, symmetric adjacency relations $\phi
  \subseteq D \times D$, and families of functions $\{ f_i : D \times A \to
  \Dist(A) \}_{i < n}$, the following implication holds: if for every initial
  value $a \in A$ and $i < n$, $f_i(-,a) : D \to \Dist(A)$ is $(\epsilon_i,
  \delta_i)$-differentially private w.r.t.\ $\phi$, then $F(-,a)$ is
  $(\epsilon^*, \delta^*)$-differentially private w.r.t.\ $\phi$ and any initial
  value $a \in A$ where \mbox{$F : (d, a) \mapsto ( \bigcirc_{i < n}
  \lift{(f_i(d,-))})(\dunit{a})$} is the $n$-fold composition of the functions
  $[f_i]_{i<n}$ and $(\epsilon^*, \delta^*) \eqdef r([(\epsilon_i, \delta_i)]_{i
  < n})$.
\end{defi}

We have the following reduction, a symmetric version of Lemma~\ref{l:star:comp}.

\begin{lem} \label{l:symstar:comp}
  Let $r : (\RR^{\geq 0}_2)^* \to \RR^{\geq 0}_2$ be a symmetric DP-composition
  rule. Let $n \in \NN$ and assume given two families of sets $\{A_i\}_{i
  \leq n}$ and $\{B_i\}_{i \leq n}$, together with a family of binary relations
  $\{ \oR(i) \subseteq A_i \times B_i \}_{i \leq n}$.
  Fix two families of functions
  $\{ g_i : A_i \to \Dist(A_{i+1}) \}_{i < n}$ and
  $\{ h_i : B_i \to \Dist(B_{i+1}) \}_{i < n}$ s.t.
  for any $i < n$ and $(a, b) \in \oR(i)$ we have:
  \begin{enumerate}
  \item
    $g_i(a) \mathrel{\symliftnew{\oR(i + 1)}{\epsilon_i, \delta_i}}
    h_i(b)$ for some parameters $\epsilon_i, \delta_i \geq 0$, and
  \item
    $g_i(a)$ and $h_i(b)$ are proper distributions.
  \end{enumerate}
  Then for $(a_0, b_0) \in \oR(0)$, there exists a symmetric $\star$-lifting
  \[
    G(a_0) \mathrel{\symliftnew{\oR(n)}{\epsilon^*, \delta^*}} H(b_0) 
  \]
  where $G : A_0 \to \Dist(A_n)$ and $H : B_0 \to \Dist(B_n)$ are the $n$-fold
  compositions of $[ g_i ]_{i \leq n}$ and $[ h_i ]_{i \leq n}$
  respectively---\ie $G(a) \eqdef (\bigcirc_{i < n} \lift{g_i})(\dunit{a})$ and
  $H(b) \eqdef (\bigcirc_{i < n} \lift{h_i})(\dunit{b})$ and $(\epsilon^*,
  \delta^*) \eqdef r([(\epsilon_i, \delta_i)]_{i < n})$.
\end{lem}
\begin{proof}
  Essentially the same as the proof of Lemma~\ref{l:star:comp}. Let $D = \{
    \mathit{tt}, \mathit{ff} \}$ as before, and take $\overline{\phi} = \{
  (\mathit{tt}, \mathit{ff}), (\mathit{ff}, \mathit{tt}) \}$ be a binary
  relation on $D$.
  
  For every $i < n$ and $(a_i, b_i) \in \oR(i)$, the definition of symmetric
  $\star$-lifting gives two distributions $\mu\lside[a_i, b_i], \mu\rside[a_i,
  b_i]$ witnessing $g_i(a) \mathrel{\symliftnew{\oR(i + 1)}{\epsilon_i,
  \delta_i}}$. We define the same maps $f_i : D \times (A^\star \times B^\star)
  \to \Dist(A^\star \times B^\star)$ as before:
  \begin{align*}
    f_i(\mathit{tt}, (a_i, b_i)) &= \mu\lside[a_i, b_i] \\
    f_i(\mathit{ff}, (a_i, b_i)) &= \mu\rside[a_i, b_i] \\
    f_i(-, (a_i, \star)) &= g_i(a_i) \times \dunit{\star} \\
    f_i(-, (\star, b_i)) &= \dunit{\star} \times h_i(b_i) \\
    f_i(-, (\star, \star)) &= \dunit{(\star, \star)}
  \end{align*}
  for $(a_i, b_i) \in \oR(i)$; otherwise, $f_i(-, (a, b)) = 0$.

  Compared to proof of Lemma~\ref{l:star:comp}, the crucial difference is that
  since we have witnesses to a \emph{symmetric} approximate lifting, the
  resulting maps $f_i(-, (a, b)) : D \to \Dist(A^\star \times B^\star)$ are
  $(\epsilon_i, \delta_i)$-differentially private with respect to the
  \emph{symmetric} relation $\overline{\phi}$, not just the \emph{asymmetric}
  relation $\phi$.  Hence, we may apply the symmetric DP-composition rule $r$
  and conclude as before.
\end{proof}

With this reduction we hand, we can generalize the advanced composition
theorem from differential privacy to $\star$-liftings.

\begin{thm}[Advanced composition \citep{DRV10}] \label{t:advcomp:priv}
  Consider a \emph{symmetric} adjacency relation $\phi$ on databases $D$.  Let
  $f_i : D \times A \to \Dist(A)$ be a sequence of $n$ functions, such that for
  every $a \in A$ the functions $f_i(-,a) : D \to \Dist(A)$ are $(\epsilon,
  \delta)$-differentially private with respect to $\phi$.  Then, for every $a \in
  A$ and $\omega \in (0, 1)$, running $f_1, \dots, f_n$ in sequence is
  $(\epsilon^*, \delta^*)$-differentially private for
  \[
    \epsilon^* = \left(\sqrt{2 n \ln(1/\omega)}\right) \epsilon +
    n \epsilon(e^\epsilon - 1)
    \quad \text{and} \quad
    \delta^* = n \delta + \omega .
  \]
\end{thm}

\begin{cor} \label{c:advcomp:star}
  Let $n$ be a natural number, $\epsilon, \delta \geq 0$, and $\omega \in (0,
  1)$ be real parameters. Suppose we have:
  \begin{enumerate}
    \item sets $\{ A_i \}_i, \{ B_i \}_i$ with $i$ ranging from $0, \dots, n$;
    \item relations $\{ \oR(i) \}_i$ on $A_i$ and $B_i$ with $i$ ranging
      from $0, \dots, n$; and
    \item functions $\{ f_i : A_i \to \Dist(A_{i + 1}) \}_i, \{ g_i : B_i \to
      \Dist(B_{i + 1}) \}_i$ with $i$ ranging from $0, \dots, n - 1$
  \end{enumerate}
  such that for all $(a, b) \in \oR(i)$, we have
  \[
    f_i(a) \mathrel{\symliftnew{\oR(i + 1)}{\epsilon, \delta}} g_i(b) 
  \]
  and $f_i(a), g_i(b)$ proper distributions. Then, there is an approximate
  lifting of the compositions:
  \[
    F(a_0)
    \mathrel{\symliftnew{\oR(n)}{\epsilon', \delta'}}
    G(b_0) 
  \]
  for every $(a_0, b_0) \in \oR(0)$, where $F : A_0 \to \Dist(A_n)$ and $G : B_0
  \to \Dist(B_n)$ are the $n$-fold (Kleisli) compositions of $\{ f_i \}$ and $\{
  g_i \}$ respectively, and the lifting parameters are:
  \[
    \epsilon' \eqdef \epsilon \sqrt{2 n \ln(1/\omega)} + n \epsilon (e^\epsilon - 1)
    \quad\quad
    \delta' \eqdef n \delta + \omega .
  \]
\end{cor}
\begin{proof}
  By the advanced composition theorem for differential privacy
  (Theorem~\ref{t:advcomp:priv}), the map $r([(\epsilon, \delta)]_{i < n})
  \eqdef (\epsilon', \delta')$ is a symmetric DP composition rule. So, we can
  conclude by Lemma~\ref{l:symstar:comp}.
\end{proof}

\section{%
  \texorpdfstring
  {$\star$-Lifting for $f$-Divergences}
{*-Lifting for f-Divergences}} \label{s:fdiv}

The definition of $\star$-lifting can be extended to lifting
constructions based on general $f$-divergences, as previously proposed
by \citet{BartheO13,OlmedoThesis}. Roughly, a $f$-divergence is a function $\Delta_f
(\mu_1,\mu_2)$ that measures the difference between two probability
distributions $\mu_1$ and $\mu_2$.  Much like we generalized the
definition for $(\epsilon, \delta)$-liftings, we can define
$\star$-lifting with $f$-divergences. Let us first
formally define $f$-divergences.
We denote by $\mathcal{F}$ the set of non-negative convex
functions vanishing at $1$: $\mathcal{F} = \{ f:\RR^{\geq 0} \rightarrow
\RR^{\geq 0} \mid f(1)=0\}$.  We also adopt the following notational
conventions: $0 \cdot f(0/0) \eqdef 0$, and $0 \cdot f(x/0)
\eqdef x \cdot \lim_{t \to 0^+} t \cdot f(1/t)$; we write $L_f$ for the limit.

\begin{defi}
  Given $f \in \mathcal{F}$ , the $f$-\emph{divergence} $\Delta_f (\mu_1,\mu_2)$
  between two distributions $\mu_1$ and $\mu_2$  in $\Dist(A)$ is defined as
  \[
    \Delta_f (\mu_1,\mu_2)  = 
      \sum_{a\in A} \mu_1(a)f\left( \frac{\mu_1(a)}{\mu_2(a)} \right) .
  \]  
\end{defi}

Examples of $f$-divergences include statistical distance ($f(t)= \frac{1}{2}
\left| t - 1 \right|$), Kullback-Leibler divergence ($f(t)= t \ln (t) - t +
1$),\footnote{%
  The additional term $- t + 1$ extends the classical definition of
KL-divergence to sub-distributions~\citep{BartheO13}.}
and Hellinger distance $(f(t)=\frac{1}{2} (\sqrt{t}-1)^2$).

\begin{defi}[$\star$-lifting for $f$-divergences]
  Let $\mu_1 \in \Dist(A)$ and $\mu_2 \in \Dist(B)$ be distributions,
  $\rR$ be a binary relation over
  $A$ and $B$, and $f\in\mathcal{F}$.
  An $(f; \delta)$-\emph{approximate lifting} of
  $\mu_1$ and $\mu_2$ for $\rR$ is a pair of distributions
  $\eta\lside \in \Dist(A \times B^\star)$ and
  $\eta\rside \in \Dist(A^\star \times B)$ s.t.
  \begin{itemize}
  \item $\dfst(\eta\lside) = \mu_1$ and $\dsnd(\eta\rside) = \mu_2$;
  \item
    $\supp({\eta\lside}_{|A \times B}) ,
     \supp({\eta\rside}_{|A \times B}) \subseteq \rR$; and
  \item $\Delta_{f}(
           \overline{\eta\lside},
           \overline{\eta\rside}) \leq \delta$,
  \end{itemize}
  where $\overline{\eta_{\bullet}}$ is the canonical lifting of $\eta_{\bullet}$
  to $A^\star \times B^\star$.
   We will write:
  $ \mu_1 \aliftnew{R}{f; \delta} \mu_2$
  if there exists an $(f; \delta)$-approximate lifting of
  $\mu_1$ and $\mu_2$ for $\rR$.
\end{defi}

$\star$-liftings for certain $f$-divergences compose sequentially.

\begin{lem} \label{l:flift:basics}
  Suppose $f$ corresponds to statistical distance, Kullback-Leibler, or
  Hellinger distance. For $i \in \{ 1, 2 \}$, let $\mu_i \in \Dist(A_i)$ and
  $\eta_i : A_i \to \Dist(B_i)$. Let $\rR$ (resp. $\rS$) be a binary relation
  over $A_1$ and $A_2$ (resp. over $B_1$ and $B_2$). If
  $\mu_1 \aliftnew{\rR}{f;\delta} \mu_2$
  for some $\delta \geq 0$ and for any $(a_1, a_2) \in \rR$ we have
  $\eta_1(a_1) \aliftnew{\rS}{f;\delta'} \eta_2(a_2)$
  for some $\delta' \geq 0$,
  then
  \[ \sE {\mu_1} {\eta_1}
    \aliftnew{\rS}{f; \delta + \delta'}
    \sE {\mu_2} {\eta_2} . \]
\end{lem}
\begin{proof}
  Essentially the same as the proof of Lemma~\ref{l:star:comp}, lifting known
  composition results for these $f$-divergences (namely, \citet[Proposition
  5]{BartheO13}).
\end{proof}

Much like the $\star$-liftings we saw before, $\star$-liftings for
$f$-divergences have witness distributions with support determined by the
support of $\mu_1$ and $\mu_2$ (cf.\ Lemma~\ref{l:alift:supp}).

\begin{lem} \label{l:flift:supp}
  Let $\mu_1 \in \Dist(A)$ and $\mu_2 \in \Dist(B)$ be distributions such that
  $ \mu_1 \aliftnew{R}{f; \delta} \mu_2 $ .
  Then, there are witnesses with support contained in $\supp(\mu_1)^\star \times
  \supp(\mu_2)^\star$.
\end{lem}

\proofatend
  Let $\mu\lside$ and $\mu\rside$ be any pair of witnesses to the approximate
  lifting.  We will construct witnesses $\eta\lside, \eta\rside$ with the
  desired support. For ease of notation, let $S_i \eqdef \supp(\mu_i)$ for
  $i \in \{ 1, 2 \}$.  Define:
  \[
    \begin{gathered}
    \eta\lside(a, b) =
    \begin{cases}
      \mu\lside(a, b) &: (a, b) \in S_1 \times S_2 \\
      \mu\lside[a, B^\star \setminus S_2] &: b = \star
    \end{cases} \\
    \eta\rside(a, b) =
    \begin{cases}
      \mu\rside(a, b) &: (a, b) \in S_1 \times S_2 \\
      \mu\rside[A^\star \setminus S_1, b] &: a = \star
    \end{cases}
    \end{gathered}
  \]
  Evidently, $\eta\lside$ and $\eta\rside$ have support in $S_1^\star \times
  S_2^\star$. Additionally, it is straightforward to check that
  $\dfst(\eta\lside) = \dfst(\mu\lside) = \mu_1$ and $\dsnd(\eta\rside) =
  \dsnd(\mu\rside) = \mu_2$ so $\eta\lside$ and $\eta\rside$ have the desired
  marginals.

  It only remains to check the distance condition. We can compute:
  \begin{align*}
    \Delta_f(\overline{\eta\lside}, \overline{\eta\rside}) &=
    \sum_{(a, b) \in S_1 \times S_2} \eta\rside(a, b) \cdot
    f \left( \frac{\eta\lside(a, b)}{\eta\rside(a, b)} \right) \\
    &+ \sum_{a \in S_1} \eta\rside(a, \star) \cdot
    f \left( \frac{\eta\lside(a, \star)}{\eta\rside(a, \star)} \right)
    + \sum_{b \in S_2} \eta\rside(\star, b) \cdot
    f \left( \frac{\eta\lside(\star, b)}{\eta\rside(\star, b)} \right) \\
    &=
    \sum_{(a, b) \in S_1 \times S_2} \mu\rside(a, b) \cdot
    f \left( \frac{\mu\lside(a, b)}{\mu\rside(a, b)} \right)
    + \sum_{a \in S_1} \eta\lside(a, \star) \cdot L_f
    + \sum_{b \in S_2} \eta\rside(\star, b) \cdot f (0) \\
    &=
    \sum_{(a, b) \in S_1 \times S_2} \mu\rside(a, b) \cdot
    f \left( \frac{\mu\lside(a, b)}{\mu\rside(a, b)} \right) \\
    &+ \sum_{a \in S_1} \sum_{b' \in B^\star \setminus S_2} \mu\lside(a, b') \cdot L_f
    + \sum_{b \in S_2} \sum_{a' \in A^\star \setminus S_1} \mu\rside(a', b) \cdot f (0)
  \end{align*}
  Now, note that for all $b' \in B^\star \setminus S_2$, we know $\mu\rside(a,
  b') = 0$. Similarly, for all $a' \in A^\star \setminus S_1$, we know
  $\mu\lside(a', b) = 0$. Hence, the last line is equal to
  \begin{align*}
    \Delta_f(\overline{\eta\lside}, \overline{\eta\rside}) &=
    \sum_{(a, b) \in S_1 \times S_2} \mu\rside(a, b) \cdot
    f \left( \frac{\mu\lside(a, b)}{\mu\rside(a, b)} \right) \\
    &+ \sum_{a \in S_1} \sum_{b' \in B^\star \setminus S_2} \mu\rside(a, b')
    \cdot f \left( \frac{\mu\lside(a, b')}{\mu\rside(a, b')} \right) \\
    &\quad \quad
    + \sum_{b \in S_2} \sum_{a' \in A^\star \setminus S_1} \mu\rside(a', b) \cdot
    f \left( \frac{\mu\lside(a', b)}{\mu\rside(a', b)} \right) \\
    &= \Delta_f(\overline{\mu\lside}, \overline{\mu\rside}) \leq \delta .
  \end{align*}
  Thus, $\eta\lside$ and $\eta\rside$ witness the desired $\star$-lifting.
\endproofatend

Finally, the mapping property from Lemma~\ref{l:alift:map} holds also for these
$\star$-liftings. While the proof of Lemma~\ref{l:alift:map} relies on the
equivalence for Sato's definition, there is no such equivalence (or definition)
for general $f$-divergences.  Therefore, we must work directly with the
witnesses of the approximate lifting.
\begin{lem} \label{l:flift:map}
  Let $\mu_1 \in \Dist(A_1)$, $\mu_2 \in \Dist(A_2)$,
  $g_1 : A_1 \to B_1$, $g_2 : A_2 \to B_2$ and $\rR$ a binary relation
  on $B_1$ and $B_2$. Let $\rS$ such that $a_1 \mathrel{\rS} a_2
    \iffdef g_1(a_1) \rR g_2(a_2)$. Then
  \[
    \dlift{g_1}(\mu_1) \aliftnew{\rR}{f; \delta} \dlift{g_2}(\mu_2)
    \iff
    \mu_1 \aliftnew{\rS}{f; \delta} \mu_2 .
  \]
\end{lem}

\proofatend
  For the reverse direction, take the witnesses $\mu\lside, \mu\rside \in \Dist(A^\star
  \times A^\star)$ and define witnesses $\nu\lside \eqdef \lift{(g_1^\star
    \times g_2^\star)}(\mu\lside)$ and $\nu\rside \eqdef \lift{(g_1^\star \times
    g_2^\star)}(\mu\rside)$, where $g_1^\star \times g_2^\star$ takes a pair $(a_1,
  a_2)$ to the pair $(g_1(a_1), g_2(a_2))$ and maps $\star$ to $\star$.  The
  support and marginal requirements are clear. The only thing to check is the
  distance condition, but this follows from monotonicity of
  $f$-divergences---under the mapping $g_1^\star \times g_2^\star : A^\star
  \times A^\star \to B^\star \times B^\star$, the $f$-divergence can only
  decrease (see, \eg, \citet{csiszar2004information}).

  For the forward direction, let $\nu\lside, \nu\rside \in \Dist(B^\star \times
  B^\star)$ be the witnesses to the second lifting. By Lemma~\ref{l:flift:supp}, we
  may assume without loss of generality that $\supp(\nu\lside)$ and
  $\supp(\nu\rside)$
  are contained in
  \[
    \supp(\lift{g_1}(\mu_1))^\star \times \supp(\lift{g_2}(\mu_2))^\star
    \subseteq g_1(A)^\star \times g_2(A)^\star .
  \]
  We aim to construct a pair
  of witnesses $\mu\lside, \mu\rside \in \Dist(A^\star \times A^\star)$ to the first
  lifting. The basic idea is to define $\mu\lside$ and $\mu\rside$ based on equivalence
  classes of elements in $A$ mapping to a particular $b \in B$, and then smooth
  out the probabilities within each equivalence class.
  To begin, for $a \in A$, define
  $[a]_{g} \eqdef g^{-1}(g(a))$ and $\alpha_i(a) \eqdef \textstyle
  \Pr_{\mu_i} [\{ a \} \mid [a]_{g_i}]$.
  We take $\alpha_i(a) = 0$ when $\mu_i([a]_{f_i}) = 0$, and we let
  $\alpha_i(\star) = 0$. We define $\mu\lside$ and $\mu\rside$ as
  \begin{align*}
    \mu\lside &: (a_1, a_2) \mapsto \alpha\lside(a_1, a_2) \cdot \nu\lside(g_1(a_1), g_2(a_2)) \\
    \mu\rside &: (a_1, a_2) \mapsto \alpha\rside(a_1, a_2) \cdot \nu\rside(g_1(a_1), g_2(a_2))
  \end{align*}
  where
  \[
    \alpha\lside(a_1, a_2) =
    \begin{cases}
      \alpha_1(a_1) \cdot \alpha_2(a_2) &: a_2 \neq \star \\
      \alpha_1(a_1) &: a_2 = \star ,
    \end{cases}
    \quad\quad
    \alpha\rside(a_1, a_2) =
    \begin{cases}
      \alpha_1(a_1) \cdot \alpha_2(a_2) &: a_1 \neq \star \\
      \alpha_2(a_2) &: a_1 = \star .
    \end{cases}
  \]
  The support and marginal conditions follow from the support and marginal
  conditions of $\nu\lside$, $\nu\rside$. For instance:
  \begin{align*}
    \sum_{a_2 \in A^\star} \mu\lside(a_1, a_2)
    &= \sum_{a_2 \in A^\star} \alpha\lside(a_1, a_2) \nu\lside(g_1(a_1), g_2(a_2)) \\
    &= \alpha_1(a_1) \nu\lside(g_1(a_1), \star)
    + \sum_{a_2 \in A}
    \alpha_1(a_1) \alpha_2(a_2) \nu\lside(g_1(a_1), g_2(a_2)) \\
    &= \alpha_1(a_1) \left( \nu\lside(g_1(a_1), \star)
      + \sum_{b_2 \in g_2(A)} \nu\lside(g_1(a_1), b_2) \sum_{a_2 \in g_2^{-1}(b_2)}
    \alpha_2(a_2) \right) \\
    &= \alpha_1(a_1) \sum_{b_2 \in B^\star} \nu\lside(g_1(a_1), b_2)
    = \alpha_1(a_1) \mu_1([a_1]_{g_1})
    = \mu_1(a_1) .
  \end{align*}
  In the last line, we replace the sum over $b_2 \in g_2(A^\star)$ with a sum
  over $b_2 \in B^\star$; this holds since the support of $\lift{g_2}(\mu_2)$ is
  contained in $g_2(A)$, so we can assume that $\nu\lside(a, b_2) = 0$ for all $b_2$
  outside of $g_2(A^\star)$. Then, we can conclude by the marginal condition
  $\dfst(\nu\lside) = \lift{g_1}(\mu_1)$. The second marginal is similar.

  We now check the distance condition
  $\Delta_f(\overline{\mu\lside}, \overline{\mu\rside}) \leq \delta$.
  We can split the $f$-divergence into $\Delta_f(\overline{\mu\lside},
  \overline{\mu\rside}) = P_0 + P_1 + P_2 + P_3$, where
  \begin{align*}
    P_0 &\eqdef \mu\rside(\star, \star)
    \cdot f \left( \frac{\mu\lside(\star, \star)}{\mu\rside(\star, \star)}
    \right)
    &P_1 &\eqdef \sum_{(a_1, a_2) \in A \times A} \mu\rside(a_1, a_2)
    \cdot f \left( \frac{\mu\lside(a_1, a_2)}{\mu\rside(a_1, a_2)} \right) \\
    P_2 &\eqdef \sum_{a_1 \in A} \mu\rside(a_1, \star)
    \cdot f \left( \frac{\mu\lside(a_1, \star)}{\mu\rside(a_1, \star)} \right)
    &P_3 &\eqdef \sum_{a_2 \in A} \mu\rside(\star, a_2)
    \cdot f \left( \frac{\mu\lside(\star, a_2)}{\mu\rside(\star, a_2)} \right)
  \end{align*}
  We will handle each term separately. Evidently $P_0 = 0$. For $P_1$, we have
  \begin{align*}
    P_1 &= \sum_{(a_1, a_2) \in A \times A} \alpha\rside(a_1, a_2)
    \nu\rside(g_1(a_1), g_2(a_2))
    \cdot f \left( \frac{\alpha\lside(a_1, a_2) \nu\lside(g_1(a_1), g_2(a_2))}
      {\alpha\rside(a_1, a_2) \nu\rside(g_1(a_1), g_2(a_2))} \right) \\
    &=
    \sum_{(a_1, a_2) \mid S^{= 0}} \alpha\lside(a_1, a_2) \nu\lside(g_1(a_1), g_2(a_2)) \cdot L_f \\
    &+ \sum_{(a_1, a_2) \mid S^{\neq 0}}
    \alpha\rside(a_1, a_2) \nu\rside(g_1(a_1), g_2(a_2))
    \cdot f \left( \frac{\nu\lside(g_1(a_1), g_2(a_2))} {\nu\rside(g_1(a_1), g_2(a_2))} \right) 
  \end{align*}
  where the sets $S^{=0}$ and $S^{\neq 0}$ are
  \begin{align*}
  S^{=0} &\eqdef \{ (a_1, a_2) \mid \nu\rside(g_1(a_1), g_2(a_2)) = 0 \} \\
  S^{\neq 0} &\eqdef \{ (a_1, a_2) \mid \nu\rside(g_1(a_1), g_2(a_2)) \neq 0 \}.
  \end{align*}
  By further rearranging,
  \begin{align*}
    P_1 &=
    \sum_{(b_1, b_2) \in (g_1 \times g_2)(S^{=0})} \nu\lside(b_1, b_2) \cdot L_f
    \left( \sum_{a_1 \in g_1^{-1}(b_1)} \alpha_1(a_1)\right)
    \left( \sum_{a_2 \in g_1^{-1}(b_2)} \alpha_2(a_2)\right) \\
    &\quad \quad
    + \sum_{\mathclap{(b_1, b_2) \in (g_1 \times g_2)(S^{\neq 0})}} \nu\rside(b_1, b_2) \cdot
    f \left( \frac{ \nu\lside(b_1, b_2) }{ \nu\rside(b_1, b_2) } \right)
    \left( \sum_{a_1 \in g_1^{-1}(b_1)} \alpha_1(a_1)\right)
    \left( \sum_{a_2 \in g_1^{-1}(b_2)} \alpha_2(a_2)\right) \\[.5em]
    &= 
    \sum_{(b_1, b_2) \in (g_1 \times g_2)(S^{=0})} \nu\lside(b_1, b_2) \cdot L_f
    + \sum_{(b_1, b_2) \in (g_1 \times g_2)(S^{\neq 0})} \nu\rside(b_1, b_2) \cdot
    f \left( \frac{ \nu\lside(b_1, b_2) }{ \nu\rside(b_1, b_2) } \right) \\
    &= 
    \sum_{(b_1, b_2) \in (g_1 \times g_2)(A \times A)} \nu\rside(b_1, b_2) \cdot
    f \left( \frac{ \nu\lside(b_1, b_2) }{ \nu\rside(b_1, b_2) } \right) \\
    &= \sum_{(b_1, b_2) \in B \times B} \nu\rside(b_1, b_2) \cdot
    f \left( \frac{ \nu\lside(b_1, b_2) }{ \nu\rside(b_1, b_2) } \right) .
  \end{align*}
  The final equality is because without loss of generality, we can assume (by
  Lemma~\ref{l:flift:supp}) that $\nu\lside, \nu\rside$ are zero outside of the
  support of $\lift{g_1}(\mu_1)$ and $\lift{g_2}(\mu_2)$, which have support
  contained in $(g_1 \times g_2)(A \times A)$.

  The remaining two terms $P_2$ and $P_3$ are simpler to bound. For $P_2$, note
  that $\overline{\mu\rside}(a, \star) = 0$ for all $a \in A$. Thus:
  \begin{align*}
    P_2 &= \sum_{a_1 \in A} \alpha\lside(a_1, \star) \nu\lside(g_1(a_1), \star)
    \cdot L_f
    = \sum_{b_1 \in g_1(A)} \sum_{a_1 \in g_1^{-1}(b_1)} \alpha_1(a_1)
    \nu\lside(b_1, \star) \cdot L_f \\
    &= \sum_{b_1 \in g_1(A)} \nu\lside(b_1, \star) \cdot L_f
    = \sum_{b_1 \in B} \nu\lside(b_1, \star) \cdot L_f
    = \sum_{b_1 \in B} \overline{\nu\rside}(b_1, \star) \cdot
    f \left( \frac{ \nu\lside(b_1, \star) }{ \overline{\nu\rside}(b_1, \star) } \right)
  \end{align*}
  where the last equality is because $\overline{\nu\rside}(b, \star) = 0$ for
  all $b \in B$.

  Similarly for $P_3$, using $\overline{\mu\lside}(\star, a) =
  \overline{\nu\lside}(\star, b) = 0$ for all $a \in A$ and $b \in B$, we have:
  \begin{align*}
    P_3 &= \sum_{a_2 \in A} \alpha\rside(\star, a_2) \nu\rside(\star, g_2(a_2)) \cdot f(0)
    = \sum_{b_2 \in g_2(A)} \sum_{a_2 \in g_2^{-1}(b_2)} \alpha_2(a_2)
    \nu\rside(\star, b_2) \cdot f(0) \\
    &= \sum_{b_2 \in g_2(A)} \nu\rside(\star, b_2) \cdot f(0)
    = \sum_{b_2 \in B} \nu\rside(\star, b_2) \cdot f(0)
    = \sum_{b_2 \in B} \nu\rside(\star, b_2) \cdot
    f \left( \frac{ \overline{\nu\lside}(\star, b_2) }{ \nu\rside(\star, b_2) } \right) .
  \end{align*}
  Putting everything together, we conclude
  \[
    \Delta_f(\overline{\mu\lside}, \overline{\mu\rside}) =
    \Delta_f(\overline{\nu\lside}, \overline{\nu\rside}) \leq \delta
  \]
  by assumption, so $\mu\lside, \mu\rside$ witness the desired approximate
  lifting.
\endproofatend
  
\section{Conclusion}
We have proposed a new definition of approximate lifting that unifies all known
existing constructions and satisfies an approximate variant of Strassen's
theorem. Our notion is useful both to simplify the soundness proof of existing
program logics and to strengthen some of their proof rules.

Subsequent to the original publication of this work, researchers have explored
two extensions of $\star$-liftings. First, \citet{AH17} develop \emph{variable
approximate liftings}, a refinement of the $(\epsilon, 0)$-approximate liftings
where the $\epsilon$ parameter is real-valued function $A \times B \to \RR^{\geq 0}$
and the distance condition on witnesses is generalized to
\[
  \mu\lside(a, b) \leq e^{\epsilon(a, b)} \mu\rside(a, b) .
\]
In effect, the approximation parameters may vary over pairs of samples instead
of being a uniform upper bound. This refinement allows capturing more precise
approximation bounds, in some cases simplifying proofs of differential privacy.
An $(\epsilon, \delta)$ version of variable approximate lifting is currently not
known.

Second, \citet{SBGHK17} explore 2-liftings in the continuous case modeling
differential privacy and $f$-divergences, but also relaxations of differential
privacy based on R\'enyi divergences~\citep{BunS2016,MironovRDP}. These
2-liftings subsume $\star$-liftings; a continuous analogue of the approximate
Strassen theorem is strongly believed to hold but remains to be shown.

We see at least two important directions for future work. First, adapting
existing program logics (in particular, \textsf{apRHL}~\citep{BartheKOZ13}) to
use $\star$-liftings, and formalizing examples that were out of reach of
previous systems. Second, symmetric $\star$-liftings seem to be an important
notion---for instance, the advanced composition theorem of differential
privacy~\citep{DRV10} applies to these liftings---but only existential versions
of the definition are currently known. A universal definition, similar to Sato's
definition for asymmetric liftings, would give more evidence that symmetric
liftings are indeed a mathematically interesting abstraction, and also give a
more convenient route to constructing such liftings.

\section*{Acknowledgments}

We thank the anonymous reviewers for their helpful suggestions. This work was
partially supported by a grant from the NSF (TWC-1513694), a grant from the
Simons Foundation (\#360368 to Justin Hsu), and the ERC Starting Grant
ProFoundNet (\#679127).

\bibliography{header,main}

\newcommand{\SortNoop}[1]{}
\begin{thebibliography}{21}
\providecommand{\natexlab}[1]{#1}
\providecommand{\url}[1]{\texttt{#1}}
\expandafter\ifx\csname urlstyle\endcsname\relax
  \providecommand{\doi}[1]{doi: #1}\else
  \providecommand{\doi}{doi: \begingroup \urlstyle{rm}\Url}\fi

\bibitem[Aharoni et~al.(2011)Aharoni, Berger, Georgakopoulos, Perlstein, and
  Spr{\"{u}}ssel]{DBLP:journals/jct/AharoniBGPS11}
R.~Aharoni, E.~Berger, A.~Georgakopoulos, A.~Perlstein, and P.~Spr{\"{u}}ssel.
\newblock \href{http://dx.doi.org/10.1016/j.jctb.2010.08.002}{The max-flow
  min-cut theorem for countable networks}.
\newblock \emph{J. Comb. Theory, Ser. {B}}, 101\penalty0 (1):\penalty0 1--17,
  2011.

\bibitem[Albarghouthi and Hsu(2018)]{AH17}
A.~Albarghouthi and J.~Hsu.
\newblock \href{https://arxiv.org/abs/1709.05361}{Synthesizing coupling proofs
  of differential privacy}.
\newblock \emph{Proceedings of the {ACM} on Programming Languages}, 2\penalty0
  (POPL), Jan. 2018.
\newblock Appeared at {ACM} {SIGPLAN--SIGACT} {S}ymposium on {P}rinciples of
  {P}rogramming {L}anguages ({POPL}), Los Angeles, California.

\bibitem[Barthe and Olmedo(2013)]{BartheO13}
G.~Barthe and F.~Olmedo.
\newblock \href{http://certicrypt.gforge.inria.fr/2013.ICALP.pdf}{Beyond
  differential privacy: Composition theorems and relational logic for
  $f$-divergences between probabilistic programs}.
\newblock In \emph{International Colloquium on Automata, Languages and
  Programming (ICALP), Riga, Latvia}, volume 7966 of \emph{Lecture Notes in
  Computer Science}, pages 49--60. Springer-Verlag, 2013.

\bibitem[Barthe et~al.(2013)Barthe, K{\"{o}}pf, Olmedo, and
  Zanella{-}B{\'{e}}guelin]{BartheKOZ13}
G.~Barthe, B.~K{\"{o}}pf, F.~Olmedo, and S.~Zanella{-}B{\'{e}}guelin.
\newblock
  \href{http://software.imdea.org/~bkoepf/papers/toplas13.pdf}{Probabilistic
  relational reasoning for differential privacy}.
\newblock \emph{ACM Transactions on Programming Languages and Systems},
  35\penalty0 (3):\penalty0 9, 2013.

\bibitem[Barthe et~al.(2016{\natexlab{a}})Barthe, Fong, Gaboardi, Gr{\'e}goire,
  Hsu, and Strub]{BartheFGGHS16}
G.~Barthe, N.~Fong, M.~Gaboardi, B.~Gr{\'e}goire, J.~Hsu, and P.-Y. Strub.
\newblock \href{https://arxiv.org/abs/1606.07143}{Advanced probabilistic
  couplings for differential privacy}.
\newblock In \emph{{ACM} {SIGSAC} Conference on Computer and Communications
  Security ({CCS}), Vienna, Austria}, 2016{\natexlab{a}}.

\bibitem[Barthe et~al.(2016{\natexlab{b}})Barthe, Gaboardi, Gr{\'e}goire, Hsu,
  and Strub]{BartheGGHS16}
G.~Barthe, M.~Gaboardi, B.~Gr{\'e}goire, J.~Hsu, and P.-Y. Strub.
\newblock \href{http://arxiv.org/abs/1601.05047}{Proving differential privacy
  via probabilistic couplings}.
\newblock In \emph{{IEEE} {S}ymposium on {L}ogic in {C}omputer {S}cience
  ({LICS}), New York, New York}, 2016{\natexlab{b}}.

\bibitem[Barthe et~al.(2016{\natexlab{c}})Barthe, Gaboardi, Hsu, and
  Pierce]{BartheGHP16}
G.~Barthe, M.~Gaboardi, J.~Hsu, and B.~C. Pierce.
\newblock \href{http://doi.acm.org/10.1145/2893582.2893591}{Programming
  language techniques for differential privacy}.
\newblock \emph{{SIGLOG} News}, 3\penalty0 (1):\penalty0 34--53,
  2016{\natexlab{c}}.

\bibitem[Bun and Steinke(2016)]{BunS2016}
M.~Bun and T.~Steinke.
\newblock \href{http://dx.doi.org/10.1007/978-3-662-53641-4_24}{Concentrated
  differential privacy: Simplifications, extensions, and lower bounds}.
\newblock In \emph{{IACR} {T}heory of {C}ryptography {C}onference (TCC),
  Beijing, China}, volume 9985 of \emph{Lecture Notes in Computer Science},
  pages 635--658. Springer-Verlag, 2016.

\bibitem[Csisz{\'a}r and Shields(2004)]{csiszar2004information}
I.~Csisz{\'a}r and P.~C. Shields.
\newblock Information theory and statistics: A tutorial.
\newblock \emph{Foundations and Trends{\textregistered} in Communications and
  Information Theory}, 1\penalty0 (4):\penalty0 417--528, 2004.

\bibitem[Dwork et~al.(2006)Dwork, McSherry, Nissim, and Smith]{DMNS06}
C.~Dwork, F.~McSherry, K.~Nissim, and A.~Smith.
\newblock \href{http://dx.doi.org/10.1007/11681878_14}{Calibrating noise to
  sensitivity in private data analysis}.
\newblock In \emph{{IACR} {T}heory of {C}ryptography {C}onference (TCC), New
  York, New York}, pages 265--284, 2006.

\bibitem[Dwork et~al.(2010)Dwork, Rothblum, and Vadhan]{DRV10}
C.~Dwork, G.~N. Rothblum, and S.~Vadhan.
\newblock
  \href{http://research.microsoft.com/pubs/155170/dworkrv10.pdf}{Boosting and
  differential privacy}.
\newblock In \emph{{IEEE} {S}ymposium on {F}oundations of {C}omputer {S}cience
  (FOCS), Las Vegas, Nevada}, pages 51--60, 2010.

\bibitem[Hsu(2017)]{JHThesis}
J.~Hsu.
\newblock \emph{Probabilistic Couplings for Probabilistic Reasoning}.
\newblock PhD thesis, University of Pennsylvania, 2017.

\bibitem[Kleinberg and Tardos(2005)]{Kleinberg:2005:AD:1051910}
J.~Kleinberg and E.~Tardos.
\newblock \emph{Algorithm Design}.
\newblock Addison-Wesley, 2005.
\newblock ISBN 0321295358.

\bibitem[Lindvall(2002)]{Lindvall02}
T.~Lindvall.
\newblock \emph{Lectures on the coupling method}.
\newblock Courier Corporation, 2002.

\bibitem[Mironov(2017)]{MironovRDP}
I.~Mironov.
\newblock \href{https://arxiv.org/abs/1702.07476}{R\'enyi differential
  privacy}.
\newblock In \emph{{IEEE} {C}omputer {S}ecurity {F}oundations {S}ymposium
  ({CSF}), Santa Barbara, California}, pages 263--275, 2017.

\bibitem[Olmedo(2014)]{OlmedoThesis}
F.~Olmedo.
\newblock \emph{Approximate Relational Reasoning for Probabilistic Programs}.
\newblock PhD thesis, Universidad Polit\'ecnica de Madrid, 2014.

\bibitem[Sato(2016)]{Sato16}
T.~Sato.
\newblock \href{https://arxiv.org/abs/1603.01445}{Approximate relational
  {H}oare logic for continuous random samplings}.
\newblock In \emph{Conference on the Mathematical Foundations of Programming
  Semantics (MFPS), Pittsburgh, Pennsylvania}, volume 325 of \emph{Electronic
  Notes in Theoretical Computer Science}, pages 277--298. Elsevier, 2016.

\bibitem[Sato et~al.(2018)Sato, Barthe, Gaboradi, Hsu, and Katsumata]{SBGHK17}
T.~Sato, G.~Barthe, M.~Gaboradi, J.~Hsu, and S.~Katsumata.
\newblock \href{https://arxiv.org/abs/1710.09010}{Reasoning about divergences
  for relaxations of differential privacy}.
\newblock 2018.

\bibitem[Strassen(1965)]{strassen1965existence}
V.~Strassen.
\newblock \href{http://projecteuclid.org/euclid.aoms/1177700153}{The existence
  of probability measures with given marginals}.
\newblock \emph{The Annals of Mathematical Statistics}, pages 423--439, 1965.

\bibitem[Thorisson(2000)]{Thorisson00}
H.~Thorisson.
\newblock \emph{Coupling, Stationarity, and Regeneration}.
\newblock Springer-Verlag, 2000.

\bibitem[Villani(2008)]{Villani08}
C.~Villani.
\newblock \emph{Optimal transport: old and new}.
\newblock Springer-Verlag, 2008.

\end{thebibliography}

\ifappendix
\newpage\appendix

\section{Detailed Proofs}

In the proofs, we will sometimes refer to the witnesses of a $\star$-lifting.

\begin{nota}
  Let $\mu_1 \in \Dist(A)$ and $\mu_2 \in \Dist(B)$ be sub-distributions,
  $\epsilon, \delta \in \RR^+$ and $\rR$ be a binary relation over $A$ and $B$.
  If two distributions $\eta\lside \in \Dist(A \times B^\star)$ and $\eta\rside
  \in \Dist(A^\star \times B)$ are witnesses to the $\star$-lifting $\mu_1
  \mathrel{\aliftnew{\rR}{\epsilon, \delta}} \mu_2$, then we write:
  \[
    \dalifted {\eta\lside} {\eta\rside} \epsilon \delta {\rR} {\mu_1} {\mu_2} .
  \]
\end{nota}

\printproofs
\fi

\end{document}
